\documentclass[twocolumn]{aastex631}

\usepackage{mathrsfs}
\usepackage{ulem}
\usepackage{float}
\usepackage{amsmath}
\usepackage{longtable}
\usepackage[T1]{fontenc}
\usepackage{float} 
\usepackage[title]{appendix} 
\shorttitle{Calibration of Instrumental Drift in MAROON-X}
\shortauthors{Basant et al.}

\begin{document}

\title{Calibrating the Instrumental Drift in MAROON-X using an Ensemble Analysis}

\correspondingauthor{Ritvik Basant}
\email{rbasant@uchicago.edu}

\author[0000-0003-4508-2436]{Ritvik Basant}
\affiliation{Department of Astronomy \& Astrophysics, University of Chicago, Chicago, IL 60637, USA}

\author[0009-0005-1486-8374]{Tanya Das}
\affiliation{Department of Astronomy \& Astrophysics, University of Chicago, Chicago, IL 60637, USA}

\author[0000-0003-4733-6532]{Jacob L.\ Bean}
\affiliation{Department of Astronomy \& Astrophysics, University of Chicago, Chicago, IL 60637, USA}

\author[0000-0002-4671-2957]{Rafael Luque}
\affiliation{Department of Astronomy \& Astrophysics, University of Chicago, Chicago, IL 60637, USA}
\affiliation{NHFP Sagan Fellow}

\author[0000-0003-4526-3747]{Andreas Seifahrt}
\affiliation{Gemini Observatory/NSF NOIRLab, 670 N. A'ohoku Place, Hilo, HI 96720, USA}

\author[0000-0003-2404-2427]{Madison Brady}
\affiliation{Department of Astronomy \& Astrophysics, University of Chicago, Chicago, IL 60637, USA}

\author[0009-0003-1142-292X]{Nina Brown}
\affiliation{Department of Astronomy \& Astrophysics, University of Chicago, Chicago, IL 60637, USA}

\author[0000-0002-4410-4712]{Julian St{\"u}rmer}
\affiliation{Landessternwarte, Zentrum f{\"u}r Astronomie der Universität Heidelberg, K{\"o}nigstuhl 12, D-69117 Heidelberg, Germany}

\author[0000-0003-0534-6388]{David Kasper}
\affiliation{Department of Astronomy \& Astrophysics, University of Chicago, Chicago, IL 60637, USA}

\author[0000-0001-7409-5688]{Guðmundur Stefánsson} 
\affil{Anton Pannekoek Institute for Astronomy, University of Amsterdam, Science Park 904, 1098 XH Amsterdam, The Netherlands}

\begin{abstract}

MAROON-X is a state-of-the-art extreme precision radial velocity spectrograph deployed on the 8.1-meter Gemini-N telescope on Maunakea, Hawai'i. Using a stabilized Fabry-P\'erot etalon for wavelength and drift calibration, MAROON-X has achieved a short-term precision of $\sim$\,30\,cm\,s$^{-1}$. However, due to a long-term drift in the etalon (2.2\,cm\,s$^{-1}$ per day) and various interruptions of the instrument baseline over the first few years of operation, MAROON-X experiences RV offsets between observing runs several times larger than the short-term precision during any individual run, which hinders the detection of longer-period signals. In this study, we analyze RV measurements of 11 targets that either exhibit small RV scatter or have signals that can be precisely constrained using Keplerian or Gaussian Process models. Leveraging this ensemble, we calibrate MAROON-X's run offsets for data collected between September 2020 and early January 2024 to a precision of $\sim$0.5\,m\,s$^{-1}$. When applying these calibrated offsets to HD 3651, a quiet star, we obtain residual velocities with an RMS of $<$70\,cm\,s$^{-1}$ in both the Red and Blue channels of MAROON-X over a baseline of 29 months. We also demonstrate the sensitivity of MAROON-X data calibrated with these offsets through a series of injection-recovery tests. Based on our findings, MAROON-X is capable of detecting sub m\,s$^{-1}$ signals out to periods of more than 1,000 days.

\end{abstract}




\section{Introduction} \label{sec:intro}

The radial velocity technique remains one of the most effective methods for detecting and mapping the orbits of exoplanets. Since the Doppler signal is inversely proportional to the mass of the star, M dwarfs are particularly favorable targets for exoplanet RV searches. The low luminosity of M dwarfs also positions their habitable zones closer to the star, making it easier to detect temperate Earth analogs. The high frequency of M dwarfs in the solar neighborhood \citep{2021A&A...650A.201R} and the potential for these nearby systems to be prime targets for future direct imaging instruments, such as ELT-PCS \citep{2021Msngr.182...38K}, RISTRETTO \citep{2022SPIE12184E..1QL}, and ELT-ANDES \citep{2023arXiv231117075P} has prompted multiple campaigns focused on understanding the exoplanet population in our vicinity. {Statistical studies of planets discovered through both the transit method \citep{2015ApJ...807...45D, 2015ApJ...798..112M} and the radial velocity method \citep{2023A&A...670A.139R} suggest that nearly all M dwarfs host planets. However, only a small fraction of the expected planets around M dwarfs in the solar neighborhood have been detected.}

MAROON-X \citep{2016SPIE.9908E..18S, 2018SPIE10702E..6DS, 2020SPIE11447E..1FS, 2022SPIE12184E..1GS} was developed to advance the search for Earth analogs orbiting within the habitable zones of M dwarfs. It is a high-resolution ($R\approx85,000$) Extreme Precision Radial Velocity echelle spectrograph deployed on the $8.1$-meter Gemini-N telescope. MAROON-X features two channels with separate cross-dispersers, cameras, and detectors that simultaneously capture the spectrum with a slight overlap in wavelength: the visible or ``Blue'' channel covers wavelengths from $500 - 670$ nm, while the near-infrared (NIR) or the ``Red'' channel covers the $650 - 900$ nm range.  MAROON-X received its first light in September 2019, followed by science verification in December 2019, and has been carrying out regular science operations since May 2020. MAROON-X is operated as part of the Gemini queue but was scheduled in campaign mode with distinct 1 -- 6-week observing runs from first light until January 2024. Since February 2024, MAROON-X has been used nearly continuously with only occasional breaks due to uncompleted access of the fiber injection unit to the telescope bottom instrument port.

\begin{figure*}[ht!]
    \centering
    \includegraphics[width=1\textwidth,height=1\textheight,keepaspectratio]{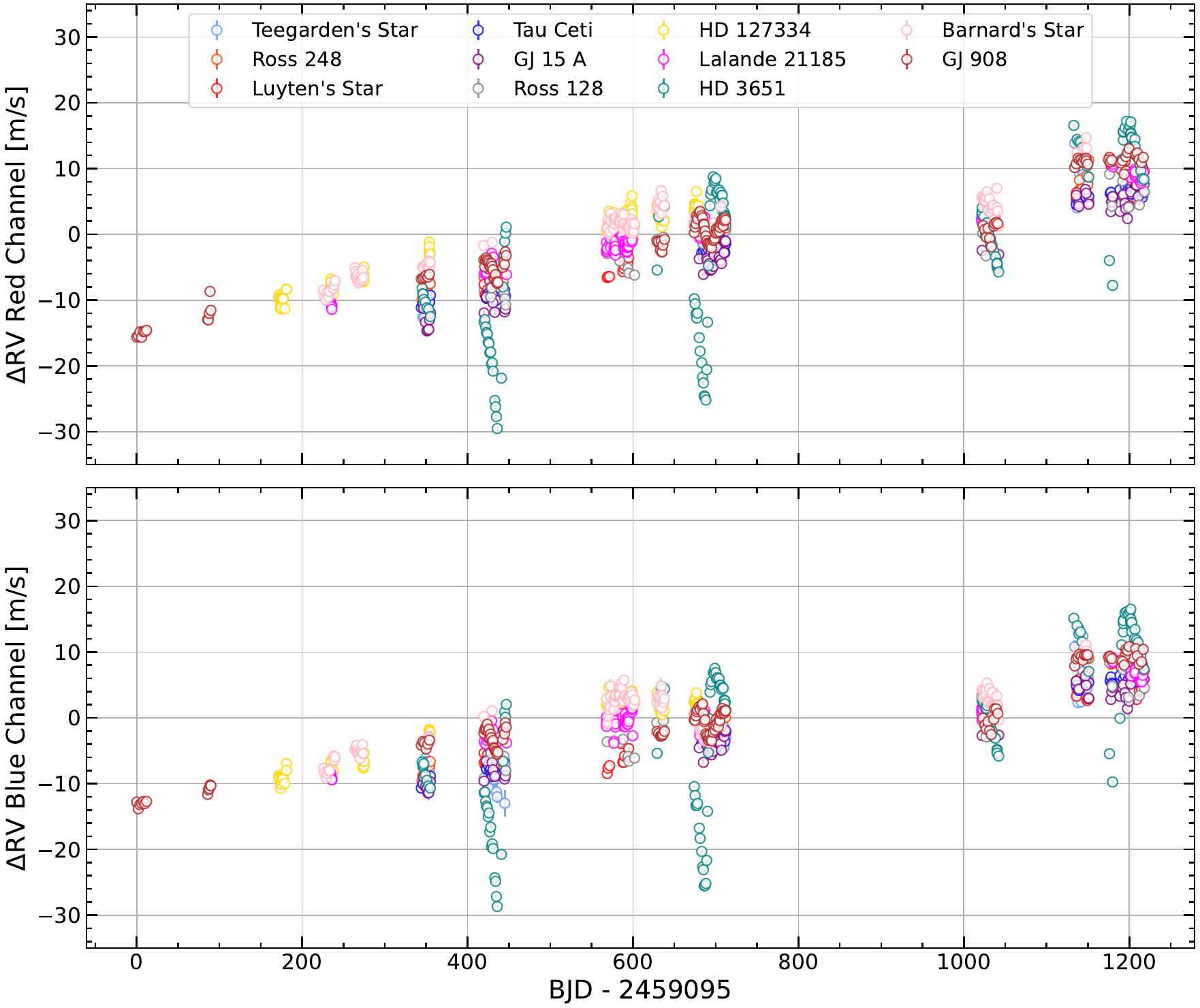}
    \caption{The differential radial velocity time series data for all the stars used in this study. The top panel displays data from the Red channel, while the bottom panel shows data from the Blue channel.}
    \label{fig:input_data}
\end{figure*}

MAROON-X data are wavelength calibrated using Fabry-P\'erot etalon spectra \citep{2017JATIS...3b5003S} that are observed simultaneously with the science targets through a dedicated fiber. The etalon was initially calibrated using a Thorium-Argon (ThAr) lamp. While the etalon spectra correct for the spectrograph's velocity variations with time (on the order of tens of m\,s$^{-1}$ nightly and seasonally), the etalon itself exhibits a long-term drift of approximately 2\,cm\,s$^{-1}$ per day (discussed in more detail in \S\ref{sec:etalon}). ThAr spectra were regularly obtained over the first few years to track this drift but were previously not utilized in the data reduction. Additionally, there have been changes in the instrumental profile due to discrete opto-mechanical perturbances caused by events such as earthquakes and power losses. The combined effect of these changes in the spectrograph and its calibration system manifests as run-to-run offsets in the MAROON-X RV time series during the campaign mode era. To mitigate these effects, one common strategy is to observe quiet stars with low RV scatter and calibrate the instrumental zero points based on on-sky data \citep[see e.g.,][]{1999ASPC..185..367U, 1999ASPC..185..354S}. This calibration method has been applied successfully to several instruments, including SOPHIE \citep{2015A&A...581A..38C}, CARMENES \citep{2018A&A...609A.117T}, HIRES \citep{2019MNRAS.484L...8T}, and HARPS \citep{2020A&A...636A..74T}.

In this work, we utilize MAROON-X data collected between September 2020 and January 2024 to calibrate the run-to-run offsets. We analyze several quiet stars, both single and multi-planetary systems, and stars whose stellar activity is well-understood to calibrate instrumental run offsets. We organized this paper as follows. In \S\ref{sec:observations_data_reduction}, we describe the MAROON-X observations and data reduction. \S\ref{sec:etalon} discusses the measurement of etalon drift using ThAr spectra. In \S\ref{sec:ensemble}, we explain our methodology for calibrating MAROON-X run offsets and briefly discuss each target used in the ensemble. Finally, in \S\ref{sec:discussion}, we compare ensemble offsets with ThAr offsets and test their sensitivity using injection-recovery tests on the radial velocity standard HD\,3651 \citep{2020AJ....160...67B} before concluding in \S\ref{sec:conclusion}.

\section{Observations and Data Reduction} \label{sec:observations_data_reduction}

For this study, we analyzed the RV time-series data of 11 targets, summarized in Table~\ref{tab:obs_data}, with a total of 861 RV measurements (in Red and Blue channels separately). The stars GJ\,908, HD\,127334, and HD\,3651 are calibration targets and were observed to calibrate the run offsets. The remaining eight targets were observed as part of a search for planets around stars of all spectral types within 4 pc. The median signal-to-noise ratio (SNR) ranges from 280 in the Red channel and 50 in the Blue channel for the faintest, late-type stars, such as Teegarden's Star, to 1300 in the Red and 940 in the Blue for the brightest, early-type stars, such as Tau Ceti. The raw RV time series output from our pipeline for all 11 targets is shown in Fig.~\ref{fig:input_data}.

The data were reduced using a custom \texttt{Python3} pipeline to produce one-dimensional, wavelength-calibrated spectra. To extract the radial velocities (RVs), we use a specific version of the \texttt{serval} package \citep{2018A&A...609A..12Z} modified to work with MAROON-X spectra. \texttt{serval} employs a template-matching approach in which the spectrum with the highest signal-to-noise ratio (SNR) is selected as a reference. All other spectra are shifted to this reference frame and co-added to create a high-SNR template. The telluric absorption and emission lines from Earth's atmosphere are completely masked to ensure they do not affect the RVs. The template is then shifted to the solar system barycenter frame to remove the contributions of Earth's motion. Finally, individual spectra are compared with the template to calculate the differential RVs via least squares fitting.  In addition, \texttt{serval} measures various activity metrics, including the chromatic index, differential line width, H$\alpha$ index, the Ca II triplet indices, and the Na I doublet index.


As mentioned earlier, MAROON-X experiences run-to-run RV offsets, which are partially introduced by the etalon calibrator itself, and partially due to the aforementioned changes in the instrumental profile of MAROON-X. Consequently, depending on the observational baseline and the time interval between consecutive runs, each target's RV time series output by \texttt{serval} may exhibit a different RV zero point. {Although both the cross-correlation function and template-matching algorithms determine differential RVs relative to a template, the absolute RVs are not relevant to the analysis presented in this work}. The only critical component is the relative change in the RVs, including the relative RV offset between different runs.

\begin{table*}[ht!]
    \centering
    \begin{tabular}{ccccccccccc}
        \hline
        \hline
         Target & Observation & Exposure Times & Baseline & No. of & \multicolumn{2}{c}{Median Error} & \multicolumn{2}{c}{Median SNR} \\
         & Semesters & [s] & [d] & Measurements & Red[\,cm\,s$^{-1}$] & Blue [\,cm\,s$^{-1}$] & Red & Blue \\
         \hline
         GJ\,15A & 2021B -- 2023B & $180$ & 871 & 62 & 31 & 30 & 620 & 318\\
         Luyten's Star & 2021B -- 2023B &  $600$ & 798  & 38 & 21 & 24 & 750 & 303\\
         HD\,3651 & 2021B -- 2023B &  $300$ & 873 & 134 & 35 & 19 & 1254 & 918\\
         Lalande\,21185 & 2021A -- 2023B & $120$ & 982 & 147 & 30 & 30 & 588 & 295 \\
         Teegarden's Star & 2021B -- 2023B & $1800$ & 804 & 34 & 31 & 121 & 280 & 46\\
         Ross\,128 & 2021B -- 2023B  & $1800$ & 790 & 33 & 20 & 26 & 750 & 260\\
         Barnard's Star & 2021A -- 2023B &  $300$ & 923 & 117 & 29 & 38 & 417 & 162\\
         GJ\,908 & 2020B -- 2023B & $300$ & 1217 & 118 & 34 & 32 & 523 & 275\\
         Tau Ceti & 2021B -- 2023B &  $45$ & 873 & 62 & 30 & 14 & 1300 & 942\\  
         Ross\,248 & 2021B -- 2023B  & $1800$ & 806 & 33 & 19 & 35 & 618 & 180\\
         HD\,127334 & 2021A -- 2022B & $300$ & 508 & 83 & 46 & 24 & 834 & 620\\
         \hline
    \end{tabular}
    \caption{This table summarizes observations for all the targets that have been used in this work.}
    \label{tab:obs_data}
\end{table*}

\section{Etalon drift measurement using ThAr calibration spectra}\label{sec:etalon}
MAROON-X uses a pressure and temperature-stabilized Fabry-P\'erot etalon, illuminated by a white-light supercontinuum laser as the main source of wavelength calibration. The etalon produces a comb of stable emission lines through constructive interference with nearly uniform spacing and intensity over a single echelle order. Over the whole spectral range of the spectrograph, both the spacing and intensity vary significantly. The wavelengths of the individual etalon lines are not known apriori and therefore, an absolute reference is required initially to determine the interference order number of each etalon line and the chromatic dispersion of the etalon. For this purpose, the spectrum of a ThAr lamp was used. While inferior in information content, particularly on a local scale, the ThAr solution is sufficient to uniquely determine the etalon parameters in conjunction with a theoretical model for the etalon dispersion  \citep{2022SPIE12184E..1GS}. {This in turn allows the subsequent calculation of an etalon-based wavelength solution for the spectrograph, independent of any other external reference \citep{ref:Bauer2015, ref:Cersullo2019}.} For simplicity, the etalon parameters for MAROON-X were determined only once, in May 2020, and presumed invariable over time, postulating a perfectly stable etalon as the wavelength reference for the spectrograph.

In reality, even a perfectly temperature- and pressure-stabilized etalon still drifts slowly over time, imprinting its changes on the spectrograph and thus compromising the long-term instrument stability. For instance, the shrinkage of the etalon spacer is known to produce achromatic linear drifts \citep{2017JATIS...3b5003S}. {Furthermore, the aging of the dielectric mirror coating of the etalon can cause chromatic drifts in the etalon, effectively changing its chromatic dispersion curve \citep{ref:Terrien2021, ref:Schmidt2022, ref:Kreider2022}}.

ThAr spectra were used to determine the etalon parameters and they can also be used to track the etalon drift. To disentangle the large drifts of the spectrograph from the small drifts of the etalon, we first wavelength-calibrate the ThAr frames using the etalon-based wavelength solution, treating the ThAr data like any other science frame. We then cross-correlate the measured ThAr spectrum order-wise against line positions derived from the NIST line list for Thorium \citep{ref:NIST2017}.
This approach has yielded narrower cross-correlation functions and smaller scatter (at the level of 20\,cm\,s$^{-1}$), both between orders and neighboring ThAr exposures since any dispersion change of the spectrograph has already been measured and corrected for using the etalon-based wavelength and drift calibration. {Since we derive each cross-correlation function from an entire echelle order, any uncorrected intra-order change of the spectrograph dispersion would otherwise lead to a spurious local RV shift of each Th line contributing to the CCF, effectively broadening and skewing the CCF if those shifts are non-uniform across an echelle order. In addition, a net shift in the centroid of the CCF is observed if all wavelengths across an echelle order have systematically drifted over time, i.e. due to a global dispersion or RV zero-point drift of the spectrograph. The chosen approach to use the etalon drift model to correct for the spectrograph drift with high granularity, i.e. including intra-order dispersion changes, enables a more accurate estimation of the etalon drift by removing the spectrograph's contribution within the limits of the chosen wavelength and drift model.} 

{The only assumption we make here is that chromatic changes of the etalon are significantly broader than an individual echelle order, as only a single RV drift value is calculated per echelle order.}

In this study, we used ThAr calibration frames that were taken over a period of two years, between August 2021 and October 2023. ThAr calibration frames taken before August 2021 were only bracketed by etalon exposures but did not use the etalon spectrum in the simultaneous calibration fiber and didn't allow for the aforementioned separation of instrument and etalon drift. Since the ThAr spectrum provides limited information content (due to non-homogeneous line distribution, blending, bleeding, and contamination with ThO), the analysis was done on a per-order basis. Echelle orders dominated by bright Ar lines or poor Th line density and observations inconsistent with neighboring frames are treated as outliers and removed from the analysis using a 3 sigma clipping method.

{Static RV offset patterns observed between the echelle orders in both channels were removed to reduce the uncertainty in the order mean (the mean over all echelle orders in a channel). This static pattern would otherwise drive up the uncertainty of the mean and possibly hide statistic evidence for the temporal variability of the etalon dispersion we are looking for here. It should be noted that the static offset pattern is likely an artifact from the selection of a different subset of Th lines used in the initial wavelength model and the subset of lines used in the cross-correlation analysis to track the change in the etalon dispersion, simply because of different rejection criteria in both steps of the analysis. Systematics in the ThAr line catalog that are affecting Th lines in the same echelle order are thus not perfectly canceling out between the wavelength solution and the cross-dispersion analysis. To a lesser extent the dispersion model of the etalon could potentially be slightly distorted by the imperfect initial ThAr-based 2D wavelength solution, further imprinting systematic errors from that solution into the dispersion model and this back into the wavelength and drift solution, which we later use to remove the drift of the spectrograph. The key is that this offset pattern is static and needs to be removed to identify temporal changes. }

\begin{figure*}[ht!]
\centering
\includegraphics[width=1\textwidth,height=1\textheight,keepaspectratio]{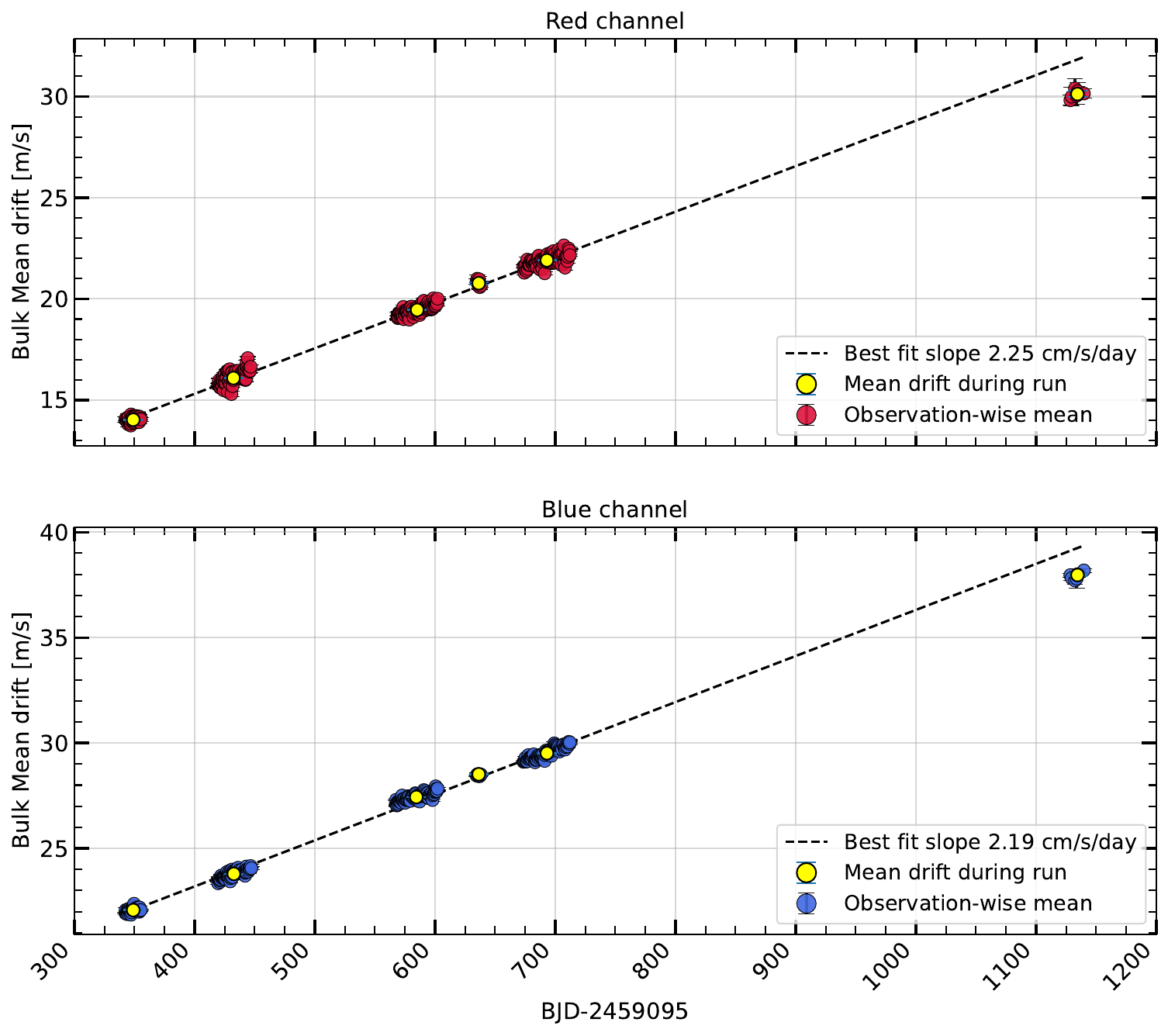}
\caption{Mean drift of the etalon from August 2021 to October 2023 as seen in Red channel (Top) and Blue channel (Bottom).}
\label{fig:alldrift}
\end{figure*}

{The data were grouped into hour-long observations to increase the SNR and the mean RV drift was computed separately for the Red and the Blue channel on a per-observation basis. We then fit a linear model to the mean RVs for each channel to estimate the long-term linear drift across all runs as well as the drift between individual runs. Our analysis shows that between August 2021 and October 2023 the Red channel drifts at a rate of  2.25\textpm0.01\,cm\,s$^{-1}$\,d$^{-1}$, while the blue channel exhibits a drift of 2.186\textpm0.008\,cm\,s$^{-1}$\,d$^{-1}$. Fig.~\ref{fig:alldrift} shows the linear drift observed in both the channels. }

\begin{figure}[]
\centering
\includegraphics[width=\columnwidth, keepaspectratio]{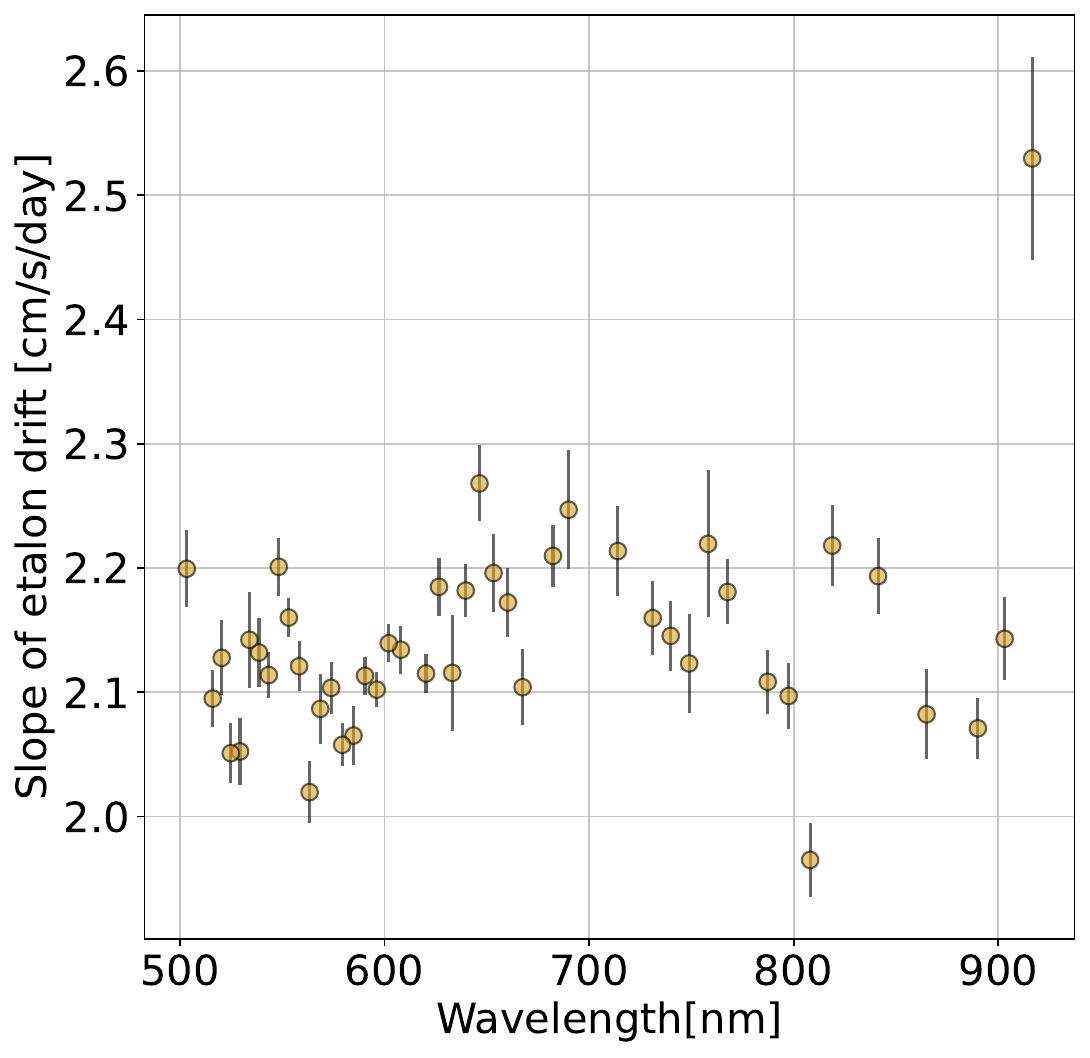}
\caption{Temporal variation in the etalon drift calculated for each order plotted as a function of wavelength.}
\label{fig:etalondrift}
\end{figure}

\begin{figure}[]
\centering
\includegraphics[width=\columnwidth, keepaspectratio]{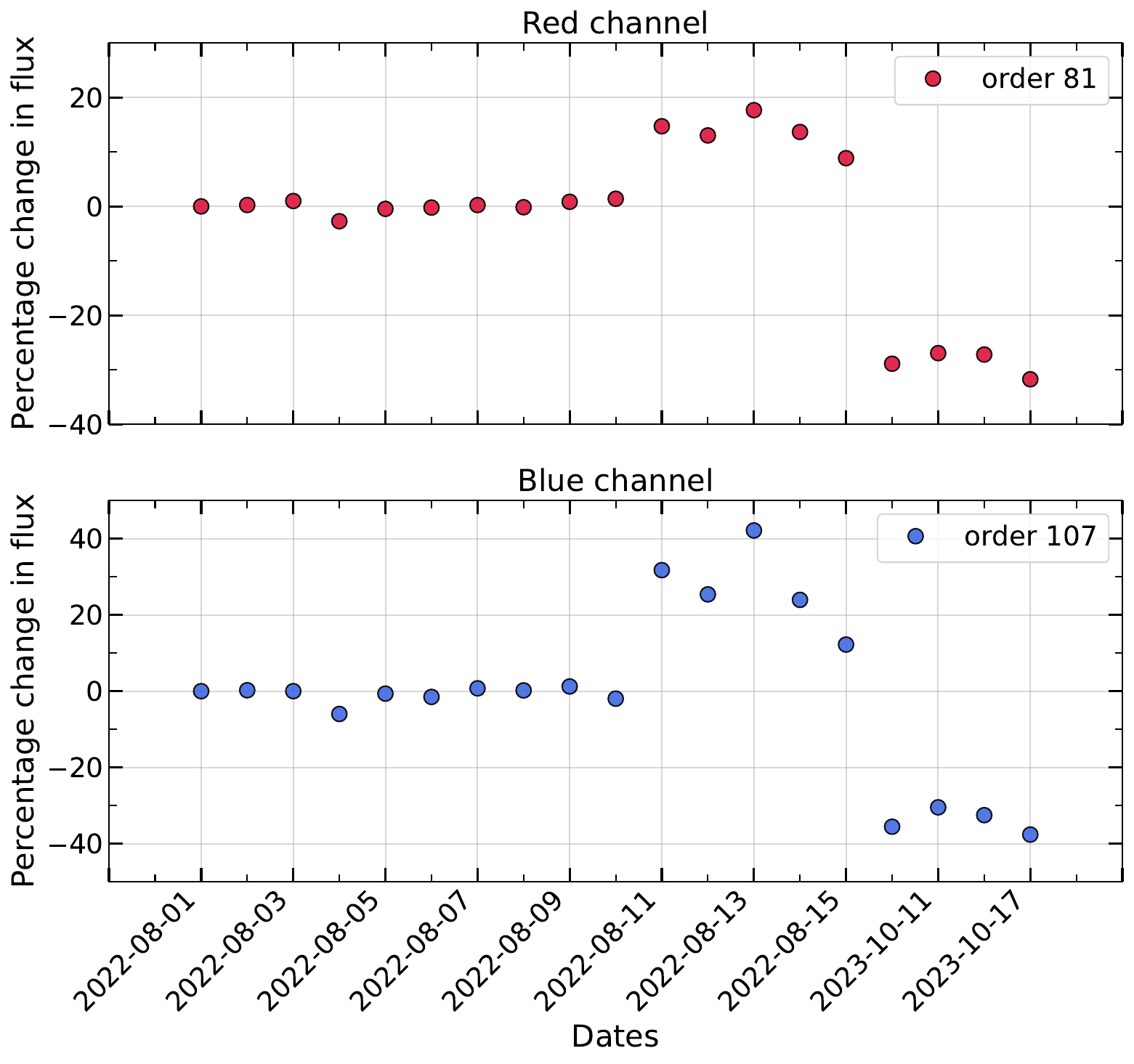}
\caption{Percentage change in flux from the ThAr lamp between August 2022 and October 2023. The change in flux is calculated with respect to the ThAr flux values on 2022-08-01 and plotted for the central order in Red channel (order 81) and blue channel (order 107) for ease of viewing.}
\label{fig:tharlamp}
\end{figure}

A linear model fitting was done across the RV values for all dates and for each individual order, and the slope of the fit is plotted with respect to the corresponding wavelength in Fig.~\ref{fig:etalondrift}. The drift rate values were estimated to be between 2.2\,cm\,s$^{-1}$\,d$^{-1}$ and 2.5\,cm\,s$^{-1}$\,d$^{-1}$ across the whole wavelength range. {The Fabry-P\'erot etalon used in HPF exhibits a wavelength dependent drift of \textpm5\,cm\,s$^{-1}$\,d$^{-1}$ \citep{ref:Terrien2021}, while that used in ESPRESSO shows a chromatic drift ranging from $-2.2$\,cm\,s$^{-1}$\,d$^{-1}$ and $+1.6$\,cm\,s$^{-1}$\,d$^{-1}$ \citep{ref:Schmidt2022}. Although the chromatic drift observed for the MAROON-X etalon is smaller in magnitude, our analysis reveals a similar pattern (see Fig.~\ref{fig:etalondrift}). This similarity is likely coincidental due to the different designs of the three Fabry-P\'erots, particularly the different mirror coatings that shape the dispersion curve.}

As evident in Fig.~\ref{fig:alldrift}, the observations around MJD 60250 (October 2023) do not conform to the trend. We find that this departure likely originates from a change in the flux of the ThAr lamp. The percentage change in flux is plotted in Fig.~\ref{fig:tharlamp}. ThAr calibrations were taken for the first fifteen days of August 2022, followed by a more than a year-long gap until October 2023. In that gap, MAROON-X was completely offline between October 2022 and June 2023 due to the Gemini North mirror accident\footnote{\url{https://noirlab.edu/public/announcements/ann22030/}}. A sudden change in the flux values can be observed around 10 August 2022 after which the flux drops significantly in October 2023, when the instrument was back online. A possible reason for these erratic changes in flux could be the aging of the ThAr lamp.

A laser frequency comb (LFC) was installed on MAROON-X in July 2023, which will be used as the absolute wavelength reference for MAROON-X. The etalon will only be used to correct the spectrograph drift in between LFC exposures, taken at the beginning and the end of a night. The LFC also provides a more precise measurement of the etalon parameters, particularly the chromatic dispersion, thereby retroactively improving the etalon-based wavelength solution of the spectrograph before the LFC became available. Likewise, the drift of the etalon can be determined much more reliably with the LFC compared to the ThAr lamp. 
Although we have been regularly collecting LFC spectra since its installation, it has not yet been incorporated into the routine data reduction pipeline. Once the LFC is integrated into the pipeline, the calibration data will be analyzed to determine instrument offsets, retroactively applying these calibrations to data collected since July 2023.

\section{Ensemble Analysis}\label{sec:ensemble}

In this section, we describe our methodology for estimating the run offsets for MAROON-X. As detailed in \S\ref{sec:intro}, MAROON-X experiences offsets between consecutive observational runs, due to changes in its instrumental profile (IP). The etalon-based wavelength and drift solution of the spectrograph does not pick up changes in the IP in the same way as \texttt{serval} does in the least-square fitting of empirical stellar templates. Etalon emission lines are unresolved and fitted with a simplified model of the IP to determine the centroid of the line in the 1D extracted spectra to calculate the spectrograph's wavelength solution. In contrast, the effective line shape of stellar spectra is a convolution of the IP and the intrinsic stellar line profile, which in most cases is marginally resolved at a Resolving power of 85,000. With different parametrization and different degrees of freedom, the response to an IP change is different between the algorithms used to compute the wavelength solution and the ones to compute the stellar RV. The result is that IP changes cause RV offsets that are not captured by the wavelength- and drift-solution of MAROON-X and that can only be derived from actual stellar observations. {We refer the readers to \citet{2024MNRAS.530.1252S} for a detailed discussion on the influence of the IP on RVs for ESPRESSO and to \citet{10.1093/pasj/psaa085} for an approach to incorporate IP changes in a forward-modeling code to determine RVs from the IRD spectrograph.}


In addition, as shown in \S\ref{sec:etalon}, the etalon calibrator itself exhibits a slow, linear drift that is small compared to the instrumental noise floor over a single observing run of a few weeks, but can accumulate to a significant offset in between runs months apart. The determination of this drift independent from stellar spectra is again limited by the impact of IP changes that affect line positions and cross-correlation peak centroids differently. 

As a result, the complete RV time series for a target observed by MAROON-X over multiple runs will exhibit a trend, with RV measurements from a given run showing noticeable shifts compared to those from prior ones.

To mitigate the impact of these drifts and ensure consistency across observational runs, it is crucial to accurately estimate the run offsets. We start by assuming that the RV measurements from different observing runs are independent and can therefore be treated as data from separate instruments. Two primary methods can be used to estimate the offsets between different runs. The simpler method involves observing a low-activity star across different runs and then using it to calibrate the instrumental shift between different observing campaigns. The second method entails modeling signals that can be precisely constrained in the RV data, while allowing the offsets between runs to vary freely. This will calibrate the offsets, which, when applied to the raw RV data, will adjust the data to allow the detection of the modeled signal. 

Although modeling known signals can help quantify instrumental changes, its effectiveness is contingent on the correct selection of signals. To address this, we begin by correcting the raw RVs by subtracting the median value for each specific observing run and clipping measurements that are more than $4\sigma$ (where $\sigma$ is the standard deviation) away from the median. We also remove data points if the RV uncertainty is greater than four times the median uncertainty. Additionally, we examine activity indices from \texttt{serval} and exclude RV measurements associated with indices that deviate substantially from their median values, which could be associated with flaring events.

We use the generalized Lomb-Scargle (GLS) periodogram \citep{2009A&A...496..577Z}, implemented via PyAstronomy \citep{2019ascl.soft06010C}, to look for the evidence of periodic signals. An analytical false alarm probability (FAP) threshold of $0.1\%$ is used to identify significant signals. It is important to note that correcting for run medians can absorb longer-period signals; therefore, we also analyze available archival data where possible.

\begin{figure*}[ht!]
    \centering
    \includegraphics[width=1.0\textwidth,height=1.0\textheight,keepaspectratio]{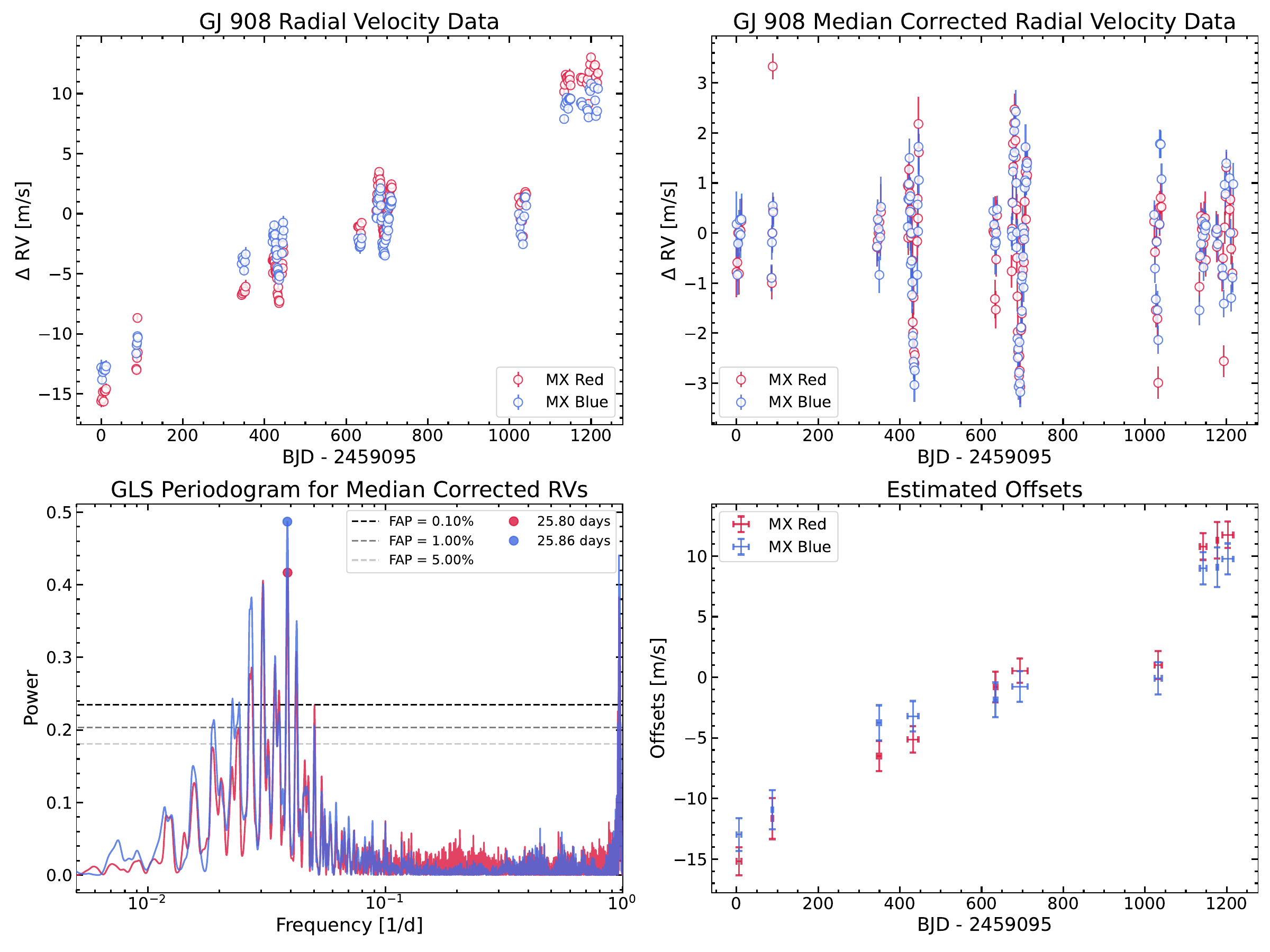}
    \caption{This figure illustrates the process of estimating offsets, with datapoints color-coded according to the channels. \textit{\textbf{Top Left}:} Raw data for GJ 908. \textit{\textbf{Top Right}:} GJ 908 data corrected by the median for individual runs. \textit{\textbf{Bottom Left}:} Computed GLS periodograms for the median-corrected data for GJ 908. A significant signal is detected at a period of $\sim 26$ d that likely corresponds to stellar rotation. \textit{\textbf{Bottom Right}:} Offsets estimated by subtracting the best fit GP model from the raw data in the top-left panel.}
    \label{fig:flowchart}
\end{figure*}

If a significant signal is identified, we fit it to the raw RV time series using the versatile modeling tool \texttt{juliet} \citep{2019MNRAS.490.2262E}, which utilizes \texttt{RadVel} \citep{2018PASP..130d4504F} and \texttt{dynesty} \citep{2020MNRAS.493.3132S} to perform nested sampling. To determine the run offsets ($\gamma$), we avoid using a jitter ($\sigma$) term in our RV fits so that all the changes in the instrument, that cause the RVs to drift, are picked up by the offsets parameter and not modeled as additional white noise. However, if known planetary parameters deviate significantly from the literature values or the GP overfits the data, we relax this assumption and use independent jitter terms for both Red and Blue channels, to account for potential wavelength-dependent variations in stellar and instrumental jitter. We also analyze the residuals for additional significant signals, which are only incorporated if they have been identified in previous studies and statistically improve the marginal evidence, thereby minimizing the risk of introducing spurious signals into the data. Lastly, when using a Keplerian signal to fit a periodic signal in the RV time series data, we parameterize our eccentricity ($e$) and argument of periastron ($\omega$) as $h = \sqrt{e}\sin\omega$ and $k = \sqrt{e}\cos\omega$ and sample values using a uniform prior between $-1$ and $1$ unless otherwise mentioned. As an example, our approach to offset calibration is demonstrated using GJ\,908 in Fig.~\ref{fig:flowchart}. 

\paragraph{\textbf{Gaussian Processes (GPs)}} GPs are an efficient way to model correlated noise in time series data \citep[see e.g.,][]{2015MNRAS.452.2269R}. To do this, GPs rely on kernel functions which are motivated by the physical processes likely responsible for the observed variations in the data. These kernel functions are defined using hyperparameters which capture the correlation observed between two points. \texttt{juliet} provides a vast range of GP kernels which are implemented either via \texttt{george} \citep{2015ITPAM..38..252A} or \texttt{celerite2} \citep{2017AJ....154..220F, 2018RNAAS...2...31F}. However, as \texttt{celerite2} provides a faster way to implement GPs, we limited ourselves to kernels that were offered within this framework: (1) Quasi-Periodic (QP) Kernel which is parameterized by $B$ (amplitude of the GP), $C$ (additive factor in the amplitude of the GP), $P_{rot}$ (period of the quasi-periodic GP), and $L$ (length scale of the exponential part of the GP); (2) Simple-Harmonic Oscillator (SHO) parameterized by $S_{0}$ (characteristic power of the oscillator), $\omega_{0}$ (characteristic frequency of the oscillator), and $Q$ (quality factor of the oscillator); and finally, (3) Double Simple Harmonic Oscillator (DSHO) which encompasses two SHO kernels centered at the first and second harmonics of the rotational period. It is parameterized by $\sigma$ (amplitude of the GP), $P_{rot}$ (period of the quasi-periodic GP), $Q_{0}$ (quality factor of the second oscillator), $f$ (fractional amplitude of secondary mode compared to the primary mode), and lastly, $dQ$ (difference between the quality factors of both oscillators). For a more detailed review, we refer the readers to  \citet{2017AJ....154..220F}.

\begin{table*}[ht!]
    \centering
    \begin{tabular}{ccccc}
    \hline
    \hline
         Star & Planetary Signals & Stellar Activity & Jitter & Archival Data \\
         \hline
         {GJ\,15A} & 11.44 d & 43.1 d & No & HIRES, APF, HARPS-N \\
         {Luyten's Star} & 4.72, 18.65 d & 406, 695 d & Yes & HARPS\\
         {HD\,3651} & 62.25 & - & No & HIRES, APF, LICK, EXPRES, NEID\\
         {Lalande\,21185} & 12.95, 3305 d & - & Yes & SOPHIE, CARMENES, APF \\
         Teegarden's Star & 4.91, 11.42, 26.13 d & 98 and 174 d & Yes & CARMENES, HPF, ESPRESSO\\
         Ross\,128 & 9.86 d & 83 d & Yes & HARPS, CARMENES\\
         Barnard's Star & 3.15 & $\sim70$, $140$ d & Yes & ESPRESSO\\
         GJ\,908 & - & 27.7 d & Yes & -\\
         {Tau Ceti} & - &  45.7 d & Yes & - \\
         {HD\,127334} & - & - & - & - \\
         Ross\, 248 & - & - & - & - \\
         \hline
    \end{tabular}
    \caption{This table summarizes the targets that have been used in the ensemble for estimating the offsets. The second and third columns describe the signals that have been modeled. Where feasible, we have also used the archival data from other instruments as mentioned in the fourth column.}
    \label{table:fitted_signals}
\end{table*}

Instead of relying on a single star for calibration, we use an ensemble of stars to enhance the accuracy of the run offsets. We selected 11 stars observed by MAROON-X either under the 4-pc sample or our calibration program based on the following criteria: (i) the star exhibits low scatter in RVs and doesn't show any strong periodic signals, (ii) planets have been detected orbiting these stars and the system has been well characterized by other studies, or (iii) {there is significant evidence of stellar activity that can be effectively modeled using sinusoids or GPs.} {
Appendix~\ref{app:signals} provides a detailed discussion of each target, while Table~\ref{table:fitted_signals} presents the final list of selected targets, along with the modeled signals and archival datasets used.}

\begin{figure*}[ht!]
    \centering
    \includegraphics[width=1.0\textwidth,height=1.0\textheight,keepaspectratio]{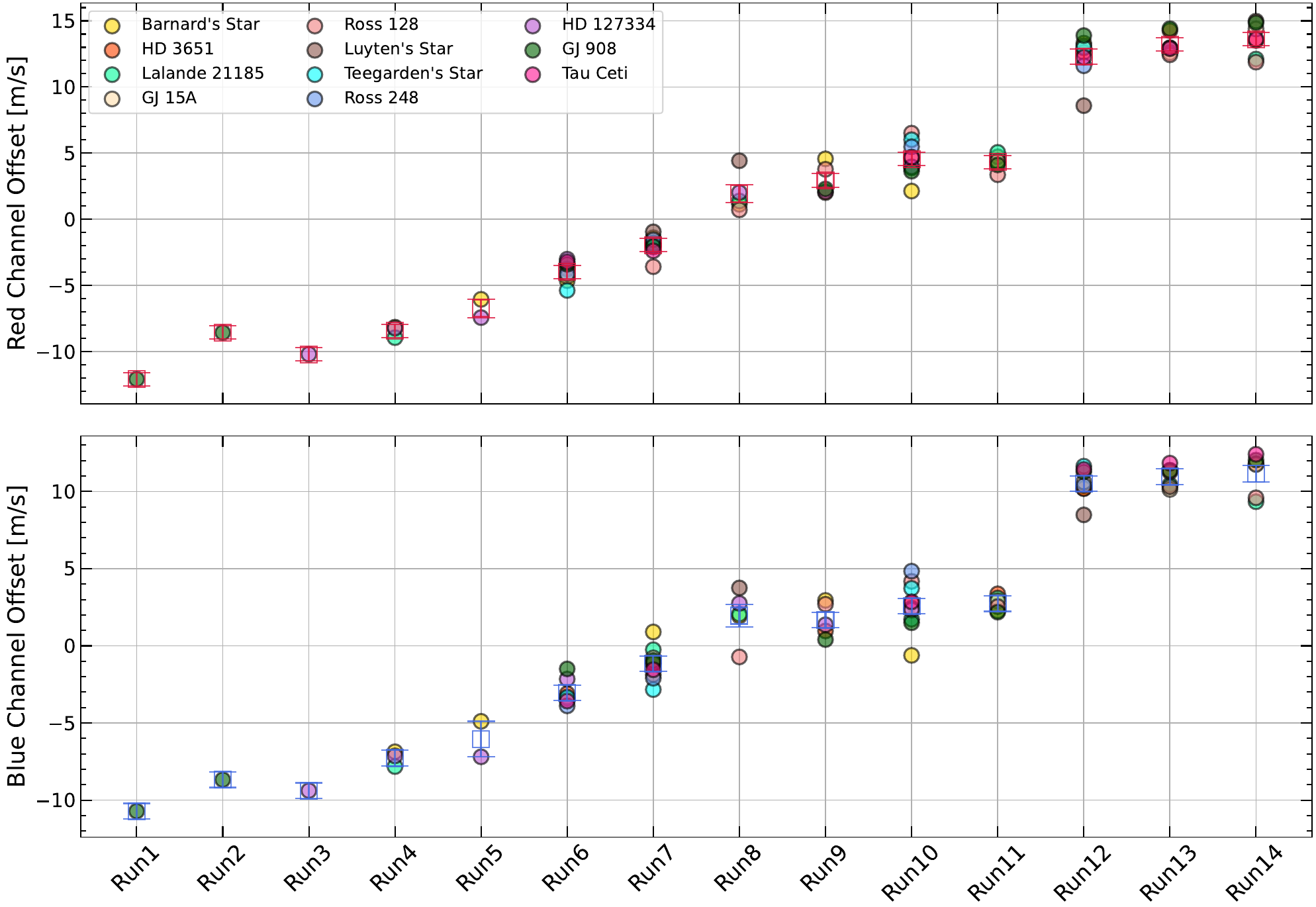} 
    \caption{The upper plot illustrates the offsets estimated using the ensemble of stars for the Red channel, while the lower plot shows the offsets for the Blue channel. The circles represent individual offsets and are color-coded by targets. The open squares indicate the median offset for all targets in a run.}
    \label{fig:Offsets}
\end{figure*}

The calibrated offsets from individual targets reflect the same range as their RV time series. However, since we used differential RVs, these offsets lack absolute meaning and are only relevant relative to one another between runs. To compute the global offset trend, we use the offsets of Barnard's star as a reference, primarily because its dataset spanned multiple observing runs. We then iterate over the subsequent stars prioritizing them based on the highest number of common runs with our reference star. To integrate a star into the ensemble, we adjust its offsets based on the average difference from the reference star’s offsets for only the common runs. Subsequently, the reference offsets are recomputed by averaging the offsets of all the stars in the ensemble. This process is repeated iteratively as additional stars are included. For all the iterations, the average difference between the reference offsets and that of the star being added is based exclusively on runs that were initially determined to be in common with Barnard's Star. This method ensures that offset values are homogenized across the ensemble while preserving the relative offsets between different runs. The offsets from individual targets are plotted in Fig.~\ref{fig:Offsets}.

\begin{table*}[ht!]
    \centering
    \begin{tabular}{cccccc}
    \hline
    \hline
         Run & Date Range & \multicolumn{2}{c}{Red Channel} & \multicolumn{2}{c}{Blue Channel} \\
          & & Offset [\,m\,s$^{-1}$] & Error [\,m\,s$^{-1}$] & Offset [\,m\,s$^{-1}$] & Error [\,m\,s$^{-1}$] \\
         \hline
         1 & 09/02/2020 - 09/14/2020 & -12.08 & 0.5    & -10.71 & 0.5    \\
         2 & 11/27/2020 - 12/01/2020 & -8.56 & 0.5    & -8.68   & 0.5   \\
         3 & 02/17/2021 - 03/03/2021 & -10.2  & 0.5   & -9.38  & 0.5   \\
         4 & 04/15/2021 - 04/29/2021 & -8.44  & 0.5 & -7.27  & 0.5 \\
         5 & 05/20/2021 - 06/03/2021 & -6.74  & 0.69 & -6.04  & 1.15 \\
         6 & 08/10/2021 - 08/22/2021 &  -3.98  & 0.5 & -3.04  & 0.5 \\
         7 & 10/26/2021 - 11/22/2021 & -1.96  & 0.5 & -1.17  & 0.5 \\
         8 & 03/23/2022 - 04/26/2022 & 1.94   & 0.66 & 1.95   & 0.74 \\
         9 & 05/23/2022 - 06/02/2022 & 2.95   & 0.52 & 1.67   & 0.5 \\
         10 & 07/07/2022 - 08/14/2022 & 4.56  & 0.5 & 2.57   & 0.5 \\
         11 & 06/20/2023 - 07/10/2023 & 4.33    & 0.5 & 2.73   & 0.5 \\
         12 & 10/09/2023 - 10/27/2023 & 12.29  & 0.58 & 10.52  & 0.5 \\
         13 & 11/20/2023 - 11/28/2023 & 13.21  & 0.5  & 10.96  & 0.5 \\
         14 & 12/04/2023 - 01/02/2024 & 13.62  & 0.5 & 11.15  & 0.54\\
         \hline
    \end{tabular}
    \caption{This table summarizes the offsets for individual runs for both the Red and Blue channels of MAROON-X.}
    \label{table:offsets}
\end{table*}

To estimate the run offsets, we compute the mean offset on a per-run basis after including all targets in the ensemble. The uncertainty in the run offsets is estimated by dividing the standard deviation of the offsets for each run by $\sqrt{N-1}$, where N is the number of targets in that run. Based on experience using these offsets for science, we adopt a lower noise floor of 0.5\,m\,s$^{-1}$. Both the mean run offsets and their uncertainties are reported in Table~\ref{table:offsets}.

\section{Discussion}\label{sec:discussion}

\subsection{Comparison with offsets from Thorium-Argon lamp}

\begin{figure}[]
\centering
\includegraphics[width=1\columnwidth, keepaspectratio]{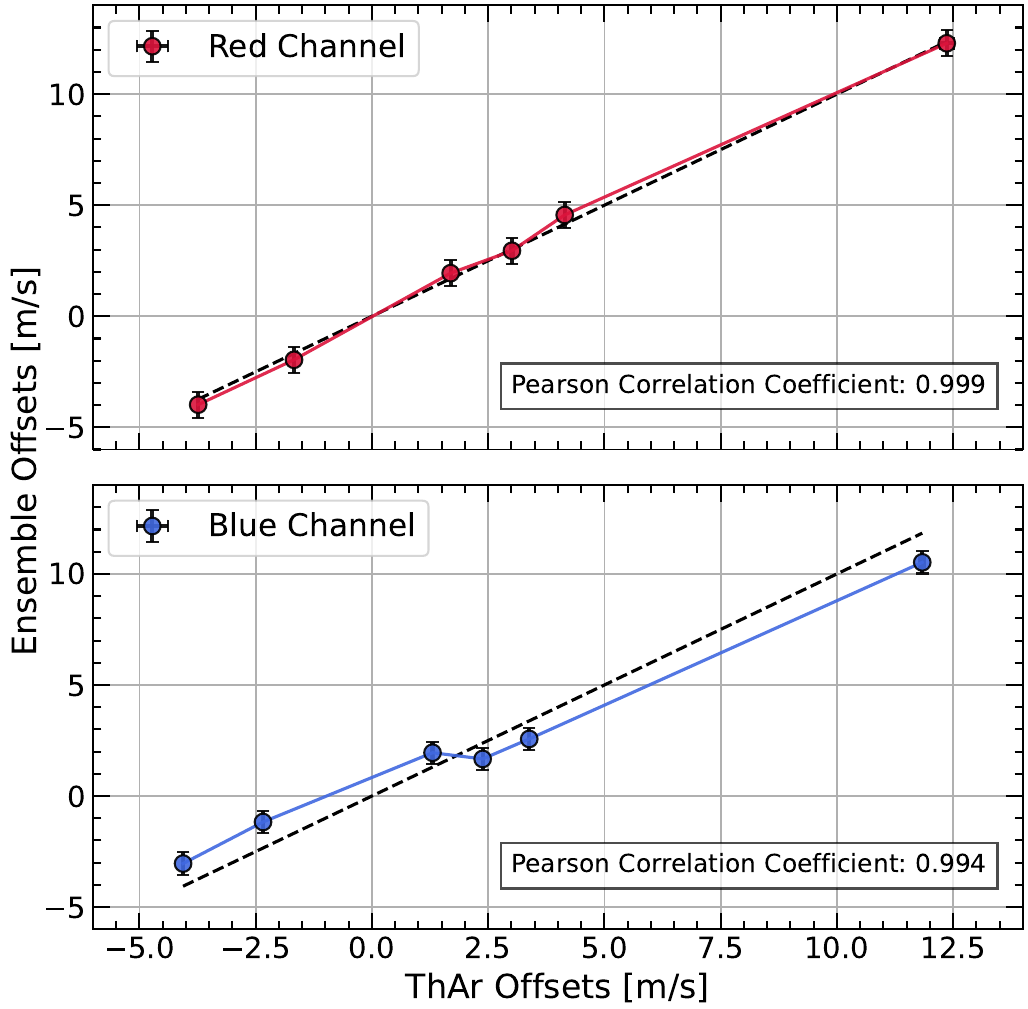}
\caption{The top panel shows the Ensemble Offsets plotted as a function of ThAr offsets for the Red channel, while the bottom panel shows it for the Blue channel.}
\label{fig:thar-offsets}
\end{figure}

The offsets derived from ThAr calibration systematically differ by a constant value from those obtained through the ensemble analysis. We quantify this difference by minimizing the root mean square (RMS) of the offsets across all available runs using \texttt{SciPy}'s minimize function. We overplot the ThAr calibration offsets with the ensemble offsets in Fig.~\ref{fig:thar-offsets}. 

{The ensemble-derived and ThAr-derived offsets are strongly correlated, with the Pearson correlation coefficient exceeding $0.99$ for both channels. To better quantify their differences, we also calculated the residuals between the two sets of offsets for a given run. For the Red channel, these differences range from $-0.28$ to $+0.41$ m\,s$^{-1}$, while for the Blue channel, they range from $-1.32$ to $1.17$ m\,s$^{-1}$. These results demonstrate that the ensemble analysis accurately tracks spectrograph variations.}

\subsection{Application Of Ensemble Offsets to HD\,3651}

We obtained 134 spectra of HD\,3651 with MAROON-X between August 13, 2021, and January 3, 2024. The typical exposure time for each visit was 5 minutes, achieving a median SNR of 1254 in the Red channel and 918 in the Blue channel. The median uncertainty in the RVs was 0.35 \,m\,s$^{-1}$ for the Red channel and 0.19 \,m\,s$^{-1}$ for the Blue channel. We applied a 4-sigma clipping to the RVs on a per-run basis, removing any data points that deviated more than four standard deviations from the median. We also excluded RVs with uncertainties greater than four times the median uncertainty. After applying these criteria, only one data point was removed from the Red dataset.

{In addition to MAROON-X observations,} we used 183 RVs for HD 3651 obtained with Keck-HIRES \citep{1994SPIE.2198..362V} between October 1996 and February 2020 \citep{2021ApJS..255....8R}. In 2004, HIRES underwent an upgrade that improved its RV precision. Therefore, we divided the HIRES dataset into two parts -- pre-upgrade and post-upgrade -- and treated them as independent instruments. We also utilized 156 RVs from NEID \citep{2016SPIE.9908E..7HS} taken between August 2021 and July 2024, accessed via the NEID archive. Due to the Contreras Fire, the WIYN Observatory was completely shut down from June 2022 to October 2022, causing an RV shift in the NEID dataset. Consequently, we treated the pre-Fire and post-Fire NEID datasets as independent instruments. Lastly, we used 61 RVs from EXPRES \citep{2016SPIE.9908E..6TJ} collected between August 2019 and February 2020 \citep{2020AJ....160...67B}. The same clipping criteria were applied to all archival data.

Since the MAROON-X offsets are defined relative to each other, we performed the orbital fitting using a modified version of \texttt{juliet} that allows for fitting relative offsets between different runs of MAROON-X data. Out of the 134 MAROON-X RVs, 32 were obtained during the 10th observing season, which spanned from July 2022 to August 2022. {We chose this run as the baseline because it includes the most observations, providing a robust reference for defining all other offsets.} An uninformative prior was used for the offsets of this baseline run. The offsets for all other runs were then defined relative to this baseline using a normal prior centered on the difference between each specific run and the reference run, with the standard deviation calculated by combining their uncertainties in quadrature. For the offsets of all other instruments -- pre-HIRES, post-HIRES, pre-NEID, post-NEID, and EXPRES -- we also used uninformative priors.

The Keplerian model was parameterized with the orbital period $P$, semi-amplitude $K$, time of periastron passage $t_{0}$, and the parameters $\text{h = }\sqrt{e}\text{sin}\omega$ and $\text{k = }\sqrt{e}\text{cos}\omega$, where $e$ is the eccentricity and $\omega$ is the argument of periastron. We used a non-informative prior for the period, semi-amplitude, and the parameters $h$ and $k$, as summarized in Table~\ref{tab:planetary-priors}. Additionally, we modeled the white noise in the RV time series using a jitter term for each instrument and for all MAROON-X runs. A log-uniform prior was chosen for all instrumental jitter terms. The posteriors and priors for derived dataset parameters are summarized in Table~\ref{tab:dataset-priors}.

\begin{table*}[]
    \centering
    \begin{tabular}{ccc}
        \hline
        \hline
        Parameter & Posterior & Prior \\
        \hline
        Orbital Period $P$ [d] & $62.249^{+0.002}_{-0.002}$ & $\mathcal{U}[60, 65]$ \\
        Semi-Amplitude $K$ [\,m\,s$^{-1}$]& $16.534^{+0.08}_{-0.077}$ & $\mathcal{U}[0, 20]$ \\
        Time of Periastron Passage $t_0$ [BJD - $2460000.5$] & $514.4^{0.2}_{0.2}$ & $\mathcal{U}[477.9, 540.9]$ \\
        $h$ & $-0.691^{+0.003}_{-0.003}$ & $\mathcal{U}[-1, 1]$ \\
        $k$ & $-0.375^{+0.005}_{-0.005}$ & $\mathcal{U}[-1, 1]$ \\
        $e$ & $0.618^{+0.003}_{-0.003}$ & - \\
        $\omega$ [ $^{\circ}$ ] & $-118.51^{+0.43}_{-0.42}$ & - \\
        \hline
    \end{tabular}
    \caption{This table summarizes the posteriors and the priors for planetary parameters used in the Keplerian model for HD 3651b.}
    \label{tab:planetary-priors}
\end{table*}

\begin{figure*}[ht!]
\centering
\includegraphics[width=1\textwidth,height=1\textheight,keepaspectratio]{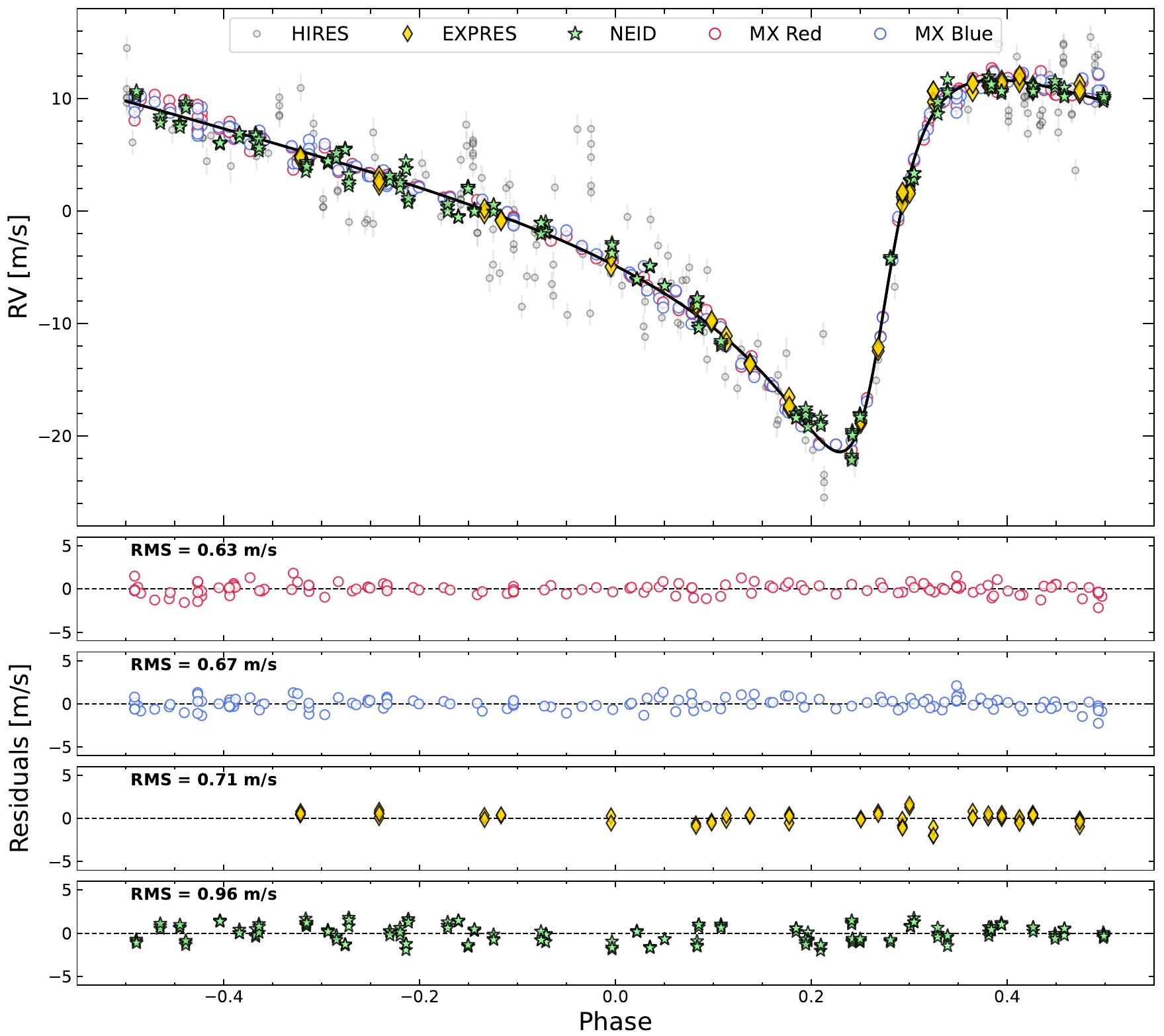}
\caption{The top panel shows the phase-folded plot for HD 3651b. A joint fit was performed between HIRES, NEID, EXPRES, and MAROON-X data. The second and third panel show the residuals for MAROON-X Red and Blue channel. The fourth and the fifth panels show the residuals for EXPRES and NEID data.}
\label{fig:HD3651_phasefolded}
\end{figure*}

The best-fit model for a single-Keplerian fit yields a period of $62.249 \pm 0.002$ days, a semi-amplitude of $16.534 \pm 0.079$ ms$^{-1}$, and an eccentricity of $0.618 \pm 0.003$. We achieve residuals with RMS values of $0.63$ ms$^{-1}$ and $0.67$ ms$^{-1}$ for the Red and Blue MAROON-X channels, respectively. These RMS values are comparable to those of other state-of-the-art instruments, such as NEID and EXPRES. The MAROON-X residuals do not show any obvious periodic signals. Fig.~\ref{fig:HD3651_phasefolded} shows the phase-folded orbit of HD 3651 b, along with the residuals from MAROON-X, EXPRES, and NEID.

\subsection{Sensitivity of Ensemble Offsets}
\begin{figure*}[ht]
    \centering
    \begin{minipage}[t]{0.48\textwidth}
        \centering
        \includegraphics[width=\textwidth,keepaspectratio]{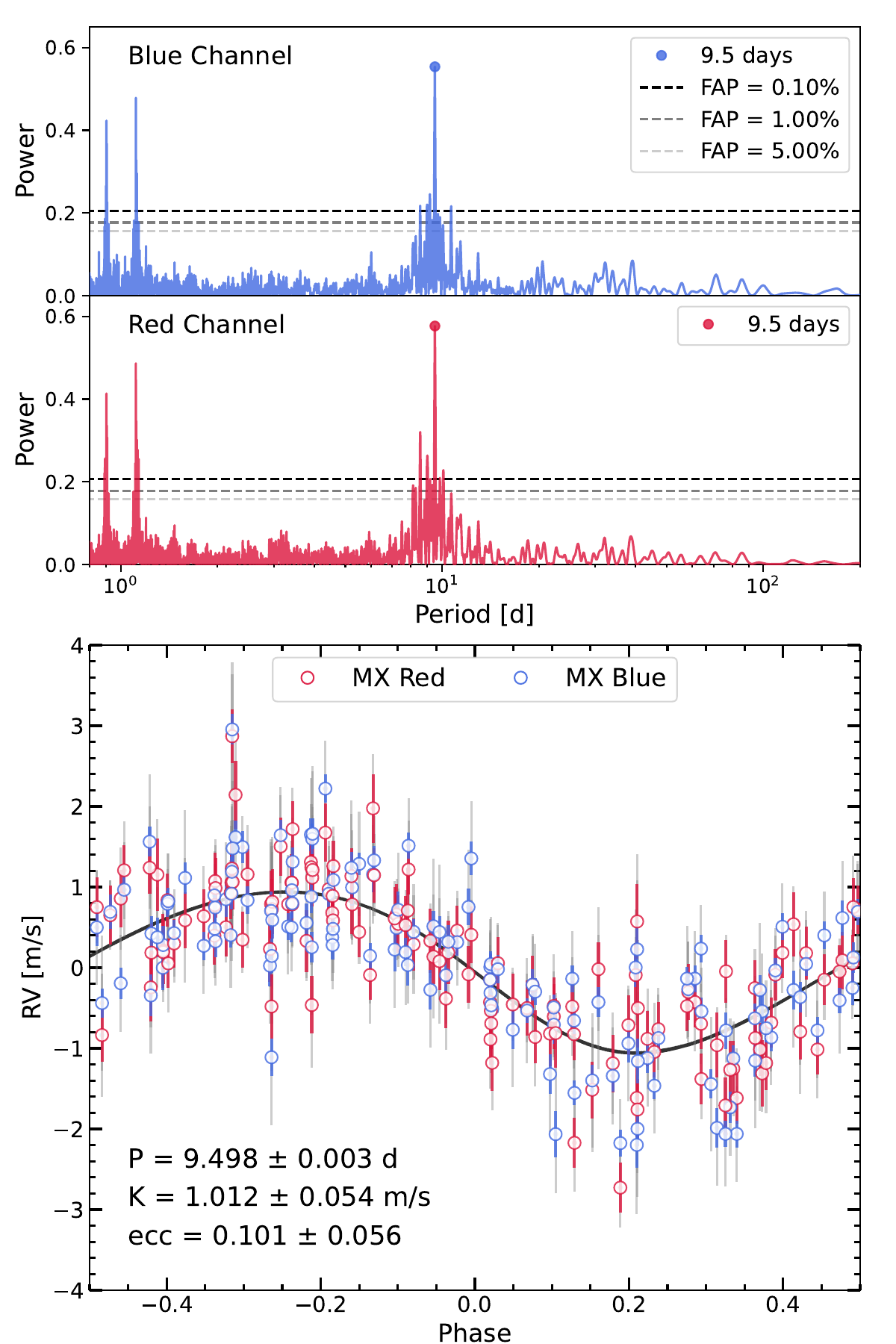}
        \caption{This figure shows results of our first injection-recovery test. A planetary signal with semi-amplitude $1$ m\,s$^{-1}$, period $9.5$ d, and eccentricity $0.1$ was injected in the raw MAROON-X RV data. The upper panel of this figure shows the GLS periodograms calculated for the residuals of MAROON-X data after removing the contribution from the known eccentric planet. The bottom panel shows the phase-folded curve for recovered planetary signal after performing a 2-Keplerian fit with the offsets from Table~\ref{table:offsets}.}
        \label{fig:injected_9.5d}
    \end{minipage}
    \hfill
    \begin{minipage}[t]{0.48\textwidth}
        \centering
        \includegraphics[width=\textwidth,keepaspectratio]{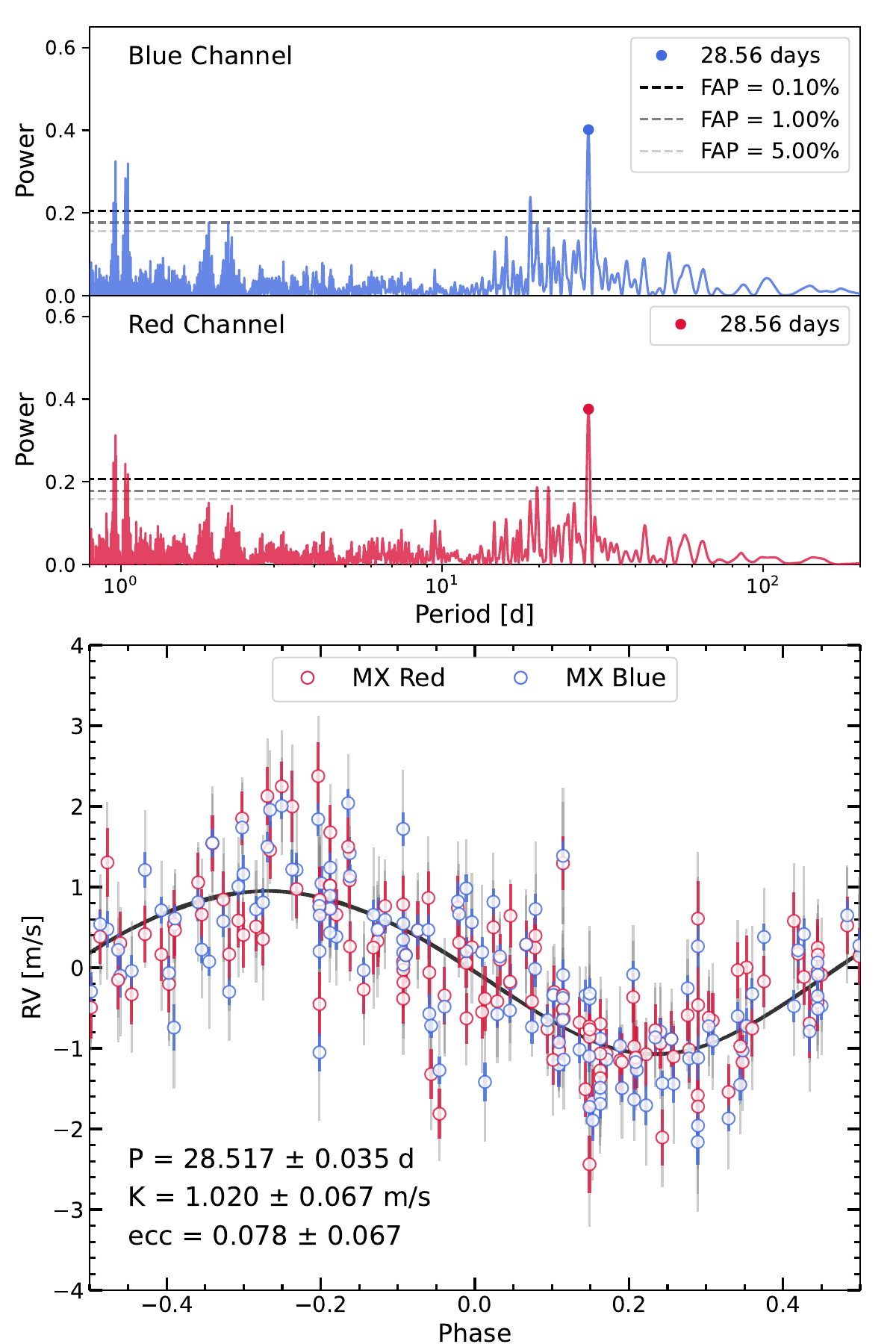}
        \caption{This figure shows results of our second injection-recovery test. A planetary signal with semi-amplitude $1$ m\,s$^{-1}$, period $28.5$ d, and eccentricity $0.1$ was injected in the raw MAROON-X RV data. The upper panel of this figure shows the GLS periodograms calculated for the residuals of MAROON-X data after removing the contribution from the known eccentric planet. The bottom panel shows the phase-folded curve for recovered planetary signal after performing a 2-Keplerian fit with the offsets from Table~\ref{table:offsets}.}
        \label{fig:injected_28.5d}
    \end{minipage}
\end{figure*}

To test the sensitivity of our ensemble offsets and their ability to detect hidden planetary signals in the RV data, we injected two identical Keplerian signals, each with a semi-amplitude of 1 \,m\,s$^{-1}$ and an eccentricity of $0.1$, into the raw RV data at periods of $9.5$ days and $28.5$ days, one at a time. After fitting for the known eccentric planet using the relative offsets, the periodograms of the residuals showed a significant peak well above the $0.1\%$ FAP at the injected periods, as illustrated in the top panels of Fig.~\ref{fig:injected_9.5d} and \ref{fig:injected_28.5d}. We then modeled the injected planet using a second Keplerian. The parameters of the recovered planets were consistent with the injected values. Both the injected planetary parameters and the recovered parameters are summarized in Table~\ref{tab:injection-recovery}. The phase-folded plots of both injected planets are shown in the lower panels of Fig.~\ref{fig:injected_9.5d} and \ref{fig:injected_28.5d}.

\begin{table}[]
    \centering
    \begin{tabular}{ccc}
        \hline
        \hline
        Parameter & Injected & Recovered \\
        \hline
        P [d] & 9.5 & $9.498^{+0.003}_{-0.003}$\\
        K [m\,s$^{-1}$] & 1.0 & $1.012^{+0.052}_{-0.055}$ \\
        e & 0.1 & $0.101^{+0.059}_{-0.058}$ \\
        \hline
        P [d] & 28.5 & $28.517^{+0.028}_{-0.038}$\\
        K [m\,s$^{-1}$] & 1.0 & $1.020^{+0.066}_{-0.068}$\\
        e & 0.1 & $0.078^{+0.088}_{-0.057}$ \\
        \hline
    \end{tabular}
    \caption{This table summarizes the planetary parameters of the injected and recovered signals.}
    \label{tab:injection-recovery}
\end{table}

\begin{figure*}[]
\centering
\includegraphics[width=1\textwidth,height=1.0\textheight,keepaspectratio]{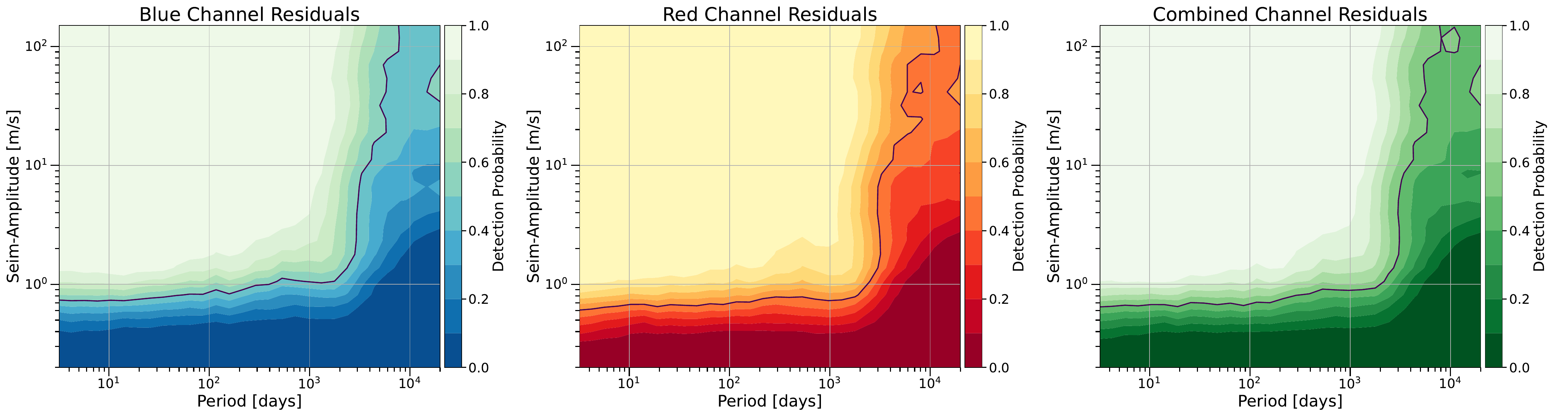}
\caption{This figure shows the completeness plots for the HD 3651~b residuals. The left panel shows the completeness for Blue data, the middle panel shows completeness for Red data, and the right panel shows completeness for combined residuals.}
\label{fig:completeness}
\end{figure*}

Additionally, we performed a suite of injection-recovery tests on the MAROON-X residuals to analyze the sensitivity of the data corrected with ensemble offsets using \texttt{rvsearch} \citep{2021ApJS..255....8R}. We generated 10,000 planetary signals with periods, semi-amplitudes, arguments of periastron, and times of periastron passage drawn from uniform distributions. The eccentricities were drawn from a beta distribution based on \citet{2013MNRAS.434L..51K}. These planetary signals were injected into the residuals, and \texttt{rvsearch} classified them as recovered if the recovered period and semi-amplitude were within $25\%$ of the injected values. A 1\,m\,s$^{-1}$ signal at a 10-day period has a detection probability of $80\%$, $85\%$, and $88\%$ in the Blue data, Red data, and combined dataset, respectively. At a 100-day period, the 1\,m\,s$^{-1}$ signal is detected with probabilities of $63\%$, $78\%$, and $79\%$ in the Blue data, Red data, and combined dataset, respectively. These sensitivity plots are shown in Fig.~\ref{fig:completeness}.

\section{Conclusions}\label{sec:conclusion}

MAROON-X is an Extreme Precision Radial Velocity spectrograph that was specifically designed to detect and characterize Earth-mass planets around M-dwarfs. However, the instrument currently suffers from a long-term drift primarily attributed to the drift in the etalon and several other operational issues that occurred in the initial years of operation.

In this paper, we attempted to characterize the instrumental drift that is observed in MAROON-X. To do this, we used an ensemble of 11 stars that were either quiet or whose RV time series showed signals that could be precisely constrained using Keplerian and Gaussian Process models. Our offsets determined using the ensemble approach are in {good} agreement {(within $<0.41$ m\,s$^{-1}$ for Red channel and within $<1.32$ m\,s$^{-1}$ for the Blue channel)} with those derived from the ThAr calibration spectra. 


We tested these ensemble offsets on HD\,3651, a quiet star with an eccentric Saturn mass planet. After applying these offsets and fitting for the known planet, the residuals from MAROON-X have an RMS that {is comparable to} other state-of-the-art instruments like NEID and EXPRES

Finally, we tested the sensitivity of our ensemble offsets by injecting planetary signals into the raw RV data on HD\,3651. We are not only able to detect the injected signals using periodograms but also recover their parameters correctly. Additionally, we also performed injection-recovery tests in the MAROON-X residuals for HD\,3651. Our analysis indicates that MAROON-X is capable of detecting sub-m\,s$^{-1}$ signals out to a period of 1,000 days. 

{While the strategy presented in this work is effective, there are a few limitations that warrant further discussion. First, although our methodology is similar to the approaches used by other instruments such as HIRES, HARPS, CARMENES, and SOPHIE to estimate the zero-point offsets, it notably diverges in the selection of stellar targets. In addition to RV-quiet stars, we incorporated stars for which RV variations can be accurately modeled. Careful evaluation of these stars is critical, as inadequate modeling may introduce systematic biases or increase offset uncertainties. Our approach also relied on a substantial dataset of over 850 RV measurements spanning more than three years, with the RVs divided into separate observational runs and the median RV subtracted to detect periodic signals. However, this median correction absorbs signals with periods longer than $\sim$ five weeks. This necessitates the inclusion of data from other instruments to ensure that all significant RV signals are accounted for. Despite these limitations, this study provides a robust method for constraining zero-point offsets, particularly in scenarios with limited observations of RV-quiet stars.}

\section*{Acknowledgements}
The University of Chicago group acknowledges funding for the MAROON-X project from the David and Lucile Packard Foundation, the Heising-Simons Foundation, the Gordon and Betty Moore Foundation, the Gemini Observatory, the NSF (award number 2108465), and NASA (grant number 80NSSC22K0117). The Gemini observations are associated with programs 21A-CAL-201, 21A-Q-119, 21A-Q-404, 21B-CAL-201, 21B-LP-202, 21B-Q-103, 21B-Q-122, 21B-Q-404, 22A-CAL-201, 22A-LP-202, 22A-Q-119, 22A-Q-409, 22B-CAL-201, 22B-LP-202, 22B-Q-119, 22B-Q-124, 22B-Q-409, 23A-CAL-201, 23A-LP-202, 23A-Q-120, 23A-Q-405, 23B-CAL-201, 23B-LP-202, 23B-Q-113, 23B-Q-405, 23B-Q-407, GN-2021B-Q-404. This paper contains data taken with the NEID instrument, which was funded by the NASA-NSF Exoplanet Observational Research (NN-EXPLORE) partnership and built by Pennsylvania State University. NEID is installed on the WIYN telescope, which is operated by the National Optical Astronomy Observatory, and the NEID archive is operated by the NASA Exoplanet Science Institute at the California Institute of Technology. NN-EXPLORE is managed by the Jet Propulsion Laboratory, California Institute of Technology under contract with the National Aeronautics and Space Administration. Support for this work was provided by NASA through the NASA Hubble Fellowship grant \#HST-HF2-51559.001-A awarded by the Space Telescope Science Institute, which is operated by the Association of Universities for Research in Astronomy, Inc., for NASA, under contract NAS5-26555.

\vspace{5mm}
\facilities{Gemini-N (MAROON-X)}

\clearpage
\appendix
\section{Modeled Signals}\label{app:signals}
\subsection{Low-Activity Stars}
\noindent Both Ross\,248 and HD\,127334 show no significant signals in median corrected periodograms. We therefore treat them as quite stars for determining the run offsets. The median of the RVs for a specific observing run is used as the offset for that run. 

\subsection{Active Stars}
\noindent Tau Ceti has been proposed to host multiple planets \citep{2017AJ....154..135F}. It is likely a pole-on star with a rotation period of $46\pm4$ d \citep{2023AJ....166..123K}. The periodogram of the median-corrected MAROON-X RVs reveal multiple significant signals between $15 - 40$ d, with the strongest being at $\sim 22$ d in the Blue channel data. This 22 d signal is close to the first harmonic of the rotational period. We attempted to model the stellar activity using a quasi-periodic GP kernel with a constraining prior on the rotational period, which is capable of effectively capturing the evolution of stellar features. To prevent saturation of MAROON-X detectors, Tau Ceti was observed consecutively 3 times with a shorter exposure time of 45 seconds on a given epoch. We binned the RVs over 15 minutes to improve the typical errors on RVs and repeated the analysis. Finally, to prevent the GP from overfitting the data, we also modeled white noise via the jitter parameter for this target. \\

\noindent GJ\,908 RVs show a remarkable change between MAROON-X observing runs, varying from extremely quiet with an RMS of $\leq$\,30\,cm\,s$^{-1}$ to showing strong periodic RV variations of several m\,s$^{-1}$ for a few observing runs. Analysis of the periodograms of the median-corrected RVs identifies the period of this signal to be $\sim26$ days. However, the analysis of the runs where these periodic variations are more prominent points to a significant signal with a period of $\sim 28$ d. Consequently, we decided to use the quasi-periodic GP kernel with a normal prior on the GP period centered at 28 days, with a standard deviation of 2 days. Binning the data with a 15-minute bin size does not change the location of periodic signals but improves the typical error on RVs. Therefore, we opted to use the binned data for our analysis. We also tested a Keplerian signal to model the RVs. Although both models produced consistent results, we opted for the GP model as it models the evolving stellar features more accurately. Additionally, the assumption of zero white noise forces the GP to overfit the data. Therefore, we also modeled white noise via the jitter term for this target. 

\subsection{Stars with Planets}
\noindent GJ\,15A is a moderately active star with a rotational period of $44$ d that hosts a short period planet \citep{2014ApJ...794...51H}. A detailed follow-up study revealed a long-term trend ($\sim 7600$ d) in the RVs \citep{2018A&A...617A.104P}. The periodogram analysis of MAROON-X median corrected RVs shows multiple significant signals between $40 - 50$ days, which coincides with the stellar rotation period. We decided to model the stellar activity via a Quasi-Periodic GP kernel. However, using an uninformative prior for the rotational period of the GP kernel converges to a longer rotational period. Fitting for the known inner planet along with the GP to model the stellar activity significantly improves the marginal evidence compared to cases where we only model either the inner planet or the stellar activity. Additionally, including the inner planet in our model results in a well-constrained rotational period of the GP that closely matches the value estimated by \citet{2014ApJ...794...51H}. As our baseline for MAROON-X data is significantly shorter than the trend, we do not include it in our analysis. Therefore, we adopt a 1-Keplerian model along with a quasi-periodic GP kernel for this target.  A more detailed analysis of this system is presented in Trifonov et al. (in prep.). Lastly, as MAROON-X observed GJ\,15A multiple times during a few nights, we decided to bin our data over 60 minutes and repeated our analysis. \\

\noindent Luyten's Star hosts two rocky planets that were discovered after modeling two additional -- more prominent -- long-period signals that likely originated from stellar activity \citep{2017A&A...602A..88A}. The periodograms for the median corrected MAROON-X RVs do not show any significant signals, possibly due to poor sampling. We therefore use archival data from HARPS to constrain the two known planets. Given our limited data on this target, we modeled the stellar activity using 2 Keplerian signals, following \citet{2017A&A...602A..88A}. The stellar rotation period for Luyten's star was estimated to be $\sim 100$ d using the R$^{'}_{HK}$ index \citep{2017A&A...600A..13A}. As the periodograms for the residuals from both the HARPS and MAROON-X data do not show any significant signals, we do not model it and adopt a 4-Keplerian model for this target. \\

\noindent HD\,3651 is a low-activity star known to host an eccentric Saturn-mass planet \citep{2003ApJ...590.1081F}. The high eccentricity of this planet renders orbits of any planet interior to it dynamically unstable \citep{2020AJ....160...67B}. This further minimizes the chances of any contamination in the RV data due to hidden short-period planets and makes it a good calibrator for our study. For this target, we adopt a 1-Keplerian model. \\

\noindent Lalande 21185 hosts a short-period rocky planet \citep{2019A&A...625A..17D}. Follow-up studies have also identified a long-term signal that may be related to stellar activity or represent an additional planetary companion \citep{2020A&A...643A.112S, 2022AJ....163..218H}. The periodogram for the median corrected MAROON-X RVs shows a significant signal at $\sim 13$ d which coincides with the known inner planet. We tested both the 1-Keplerian and 2-Keplerian models and chose to adopt the 2-Keplerian model as it provided a significant improvement in the marginal evidence. Based on the photometric observations, the stellar rotation period of Lalande 21185 has been estimated to be $56.15 \pm 0.27$ d \citep{2019A&A...625A..17D}. The periodogram for the Red channel residuals shows multiple significant signals at $\sim14$ d and between $25 - 36$ d while that for the Blue channel shows only one significant signal at $\sim 26$ d. As these signals are inconsistent between the Red and Blue channel and do not coincide with the estimated stellar rotation period, we opted not to model them. Lastly, a model with assumption of zero white noise leads to a higher eccentricity solution for the inner planet. Therefore, we decided to account for white noise for this target.\\ 

\noindent Teegarden's Star was discovered to harbor 2 short-period planets \citep{2019A&A...627A..49Z}. Recently, a follow-up study utilizing MAROON-X data found a third planet \citep{2024A&A...684A.117D} that only reveals itself after removing the two known planets and two other stellar activity signals. Although \citet{2024A&A...684A.117D} tested different GP kernels to model the stellar activity, use of Keplerian signals provided significantly higher marginal evidence. Therefore, we decided to model the stellar activity using Keplerian signals. Our eccentricity parameterization ($\sqrt{e}sin\omega$ and $\sqrt{e}cos\omega$) results in significantly higher eccentricities for the two Keplerian signals modeling the stellar activity compared to the values reported in \citet{2024A&A...684A.117D}. To avoid this, we opted to draw the eccentricities directly from a beta distribution with parameters $\alpha = 1.52$ and $\beta = 29$ and the argument of periastron from a uniform distribution $\mathcal{U}[-180, 180]$, only for the Keplerian signals modeling the stellar activity. This distribution ensures the eccentricities to remain low while still allowing for higher values to be sampled and it is a good assumption for compact multi-planetary systems with low-mass planets \citep{2019AJ....157...61V}. Thus, our final model for this target consists of 5 Keplerian signals, with the eccentricities for the signals modeling the stellar activity drawn from a beta distribution. \\

\noindent Ross 128 is known to harbor a single planet on a non-circular orbit with a period of $9.86$ d \citep{2018A&A...613A..25B}. A recent follow-up study confirmed this eccentric planet and did not find evidence for any other signal in the RVs \citep{2024A&A...690A.234L}. The periodograms of the MAROON-X median-corrected RVs do not show any evidence of periodic signals, possibly due to the limited sampling of the target. Despite this, we use the available archival data in conjunction with the MAROON-X data to fit the known planet. Following \citet{2024A&A...690A.234L}, we also tested a quasi-periodic and an SHO GP kernels to model the stellar activity. In both cases, the assumption of zero white noise yields poor convergence for the GP hyperparameters. Therefore, we chose to model white noise for this target. Although both GP kernels provide consistent results for the offsets (within $< 0.5$ m\,s$^{-1}$), we opted for the Quasi-Periodic kernel as it resulted in a significantly better marginal evidence. Our solution for the rotation period ($\sim 83.2$ d) is consistent with value estimated by \citet[][]{2024A&A...690A.234L}. Lastly, we note that our solution for the orbit of the known planet indicates a lower eccentricity ($0.09^{+0.07}_{-0.06}$).\\

\noindent Barnard's Star is a slow-rotating M3.5-type star \citep{2019MNRAS.488.5145T}. Recently, a detailed analysis utilizing ESPRESSO RVs identified a sub-Earth mass planet with a period of $3.15$ d and found evidence for 3 more candidate planets \citep{2024A&A...690A..79G}. On certain nights, MAROON-X observed Barnard's Star multiple times within the same night. We therefore decided to bin the data over 15 minutes which slightly improves the median RV error. The analysis of the periodograms for the median-corrected MAROON-X RVs reveals several significant signals between $20 - 50$ d. Fitting a Keplerian signal with broad prior on the period converges to a solution with a period of $\sim70$ d, which is close to the second harmonic of the stellar rotation period \citep{2022ApJ...927L..11T}. The periodograms of MAROON-X residuals of the 1-Keplerian model also indicate the sub-Earth planet at 3.15 d. Including the 3.15 d signal along with the 70 d signal significantly improves the marginal evidence.

The publicly available ESPRESSO data for Barnard's Star have significant overlap with the data from MAROON-X and therefore provide a more robust way to constrain the offsets. Following \citet{2024A&A...690A..79G}, we decided to use a DSHO GP kernel to model the stellar activity in the combined ESPRESSO and MAROON-X RV time series. The assumption of zero white noise leads to solutions with significantly shorter stellar rotation periods. Therefore we incorporated white noise into the model for this target. Using uninformative prior for the rotational period of the GP results in a period of $139.3$ d, consistent with the value reported by \citet{2024A&A...690A..79G}, who modeled the RVs along with the Full Width at Half Maxima measurements using a multi-dimensional GP. However, our solution doesn't converge to a unique value for the GP hyperparameters $f$ and $dQ$. A more detailed analysis of this target is out of the scope of the current work and will be presented in a different study {(Basant et al., in review)}. The use of a GP to model the stellar activity instead of a Keplerian signal yields a significant improvement in the marginal evidence. As the three candidate signals have semi-amplitudes smaller than the typical RV error provided by MAROON-X, we do not include them in our analysis. Thus, we adopt a 1-Keplerian model along with the DSHO GP kernel for this target. 

\section{MAROON-X Radial Velocity Data HD 3651}

\begin{longtable*}{ccccc}
\caption{Full MAROON-X RV dataset of HD 3651 before clipping.} \\
\hline
\hline
BJD & RV MX Red & Error MX Red & RV MX Blue & Error MX Blue \\
\text{[d]} & \text{[\,m\,s$^{-1}$]} & \text{[\,m\,s$^{-1}$]} & \text{[\,m\,s$^{-1}$]} & \text{[\,m\,s$^{-1}$]} \\
\hline
\endfirsthead
\multicolumn{5}{c}%
{{\tablename\ \thetable{} -- continued from previous page}} \\
\hline
\hline
BJD & RV MX Red & Error MX Red & RV MX Blue & Error MX Blue \\
\text{[d]} & \text{[\,m\,s$^{-1}$]} & \text{[\,m\,s$^{-1}$]} & \text{[\,m\,s$^{-1}$]} & \text{[\,m\,s$^{-1}$]} \\
\hline
\endhead
\hline
\multicolumn{5}{r}{{Continued on next page}} \\
\endfoot

\hline
\endlastfoot

2459440.128470 & -8.18 & 0.36 & -6.71 & 0.21 \\
2459441.125834 & -9.05 & 0.46 & -7.48 & 0.28 \\
2459441.129877 & -9.07 & 0.40 & -7.20 & 0.24 \\
2459442.107979 & -9.89 & 0.35 & -6.97 & 0.20 \\
2459444.006271 & -10.28 & 0.38 & -8.68 & 0.22 \\
2459445.944729 & -11.17 & 0.40 & -9.50 & 0.23 \\
2459448.067630 & -11.38 & 0.39 & -10.33 & 0.23 \\
2459449.975713 & -12.50 & 0.37 & -10.67 & 0.21 \\
2459514.997427 & -13.35 & 0.41 & -11.41 & 0.24 \\
2459515.986651 & -12.97 & 0.36 & -11.28 & 0.22 \\
2459516.967455 & -14.08 & 0.36 & -12.67 & 0.21 \\
2459517.887230 & -14.91 & 0.36 & -13.40 & 0.22 \\
2459518.933686 & -15.15 & 0.35 & -13.64 & 0.21 \\
2459520.009877 & -16.36 & 0.36 & -15.02 & 0.22 \\
2459520.910549 & -16.57 & 0.35 & -14.46 & 0.21 \\
2459521.880543 & -18.00 & 0.40 & -17.33 & 0.24 \\
2459522.936804 & -17.88 & 0.40 & -16.62 & 0.24 \\
2459523.938852 & -19.75 & 0.40 & -19.59 & 0.24 \\
2459524.882129 & -19.51 & 0.40 & -19.17 & 0.24 \\
2459525.791991 & -20.78 & 0.37 & -19.87 & 0.23 \\
2459527.926874 & -25.25 & 0.41 & -24.30 & 0.25 \\
2459528.939310 & -26.23 & 0.37 & -24.85 & 0.23 \\
2459529.904122 & -27.72 & 0.39 & -27.17 & 0.25 \\
2459530.919924 & -29.52 & 0.38 & -28.68 & 0.24 \\
2459535.955905 & -21.87 & 0.43 & -20.76 & 0.26 \\
2459537.875620 & -8.00 & 0.38 & -6.60 & 0.25 \\
2459538.961854 & -3.08 & 0.39 & -2.30 & 0.27 \\
2459539.872039 & -1.09 & 0.35 & 0.90 & 0.23 \\
2459540.945447 & 0.17 & 0.43 & 0.08 & 0.29 \\
2459541.842758 & 1.10 & 0.37 & 2.04 & 0.24 \\
2459724.113271 & -5.43 & 0.32 & -5.39 & 0.16 \\
2459726.118816 & 2.68 & 0.33 & 2.63 & 0.16 \\
2459729.081656 & 5.13 & 0.53 & 4.45 & 0.26 \\
2459730.120416 & 4.11 & 0.35 & 4.75 & 0.16 \\
2459733.092626 & 4.31 & 0.34 & 4.43 & 0.16 \\
2459769.120402 & -9.79 & 0.32 & -10.45 & 0.17 \\
2459770.110728 & -10.56 & 0.31 & -11.80 & 0.17 \\
2459771.126479 & -12.07 & 0.32 & -13.33 & 0.17 \\
2459772.121599 & -12.74 & 0.31 & -13.31 & 0.16 \\
2459773.073370 & -11.97 & 0.31 & -13.01 & 0.16 \\
2459775.095000 & -15.71 & 0.31 & -16.77 & 0.17 \\
2459776.104378 & -17.76 & 0.30 & -18.32 & 0.16 \\
2459778.106797 & -19.51 & 0.36 & -20.31 & 0.19 \\
2459779.096542 & -21.70 & 0.36 & -22.65 & 0.19 \\
2459780.108339 & -22.59 & 0.31 & -23.08 & 0.17 \\
2459781.049541 & -24.57 & 0.31 & -25.52 & 0.17 \\
2459782.117313 & -24.72 & 0.30 & -25.50 & 0.16 \\
2459783.115624 & -25.19 & 0.31 & -25.18 & 0.16 \\
2459784.092286 & -20.57 & 0.32 & -21.68 & 0.17 \\
2459785.092446 & -13.35 & 0.30 & -14.21 & 0.16 \\
2459786.077203 & -4.78 & 0.31 & -5.28 & 0.16 \\
2459787.084846 & 0.51 & 0.31 & -0.12 & 0.17 \\
2459788.095015 & 4.35 & 0.32 & 3.91 & 0.16 \\
2459789.021505 & 6.07 & 0.34 & 5.04 & 0.19 \\
2459790.041639 & 6.76 & 0.36 & 5.50 & 0.19 \\
2459792.027257 & 8.74 & 0.36 & 6.90 & 0.20 \\
2459793.050381 & 7.88 & 0.34 & 6.72 & 0.18 \\
2459793.998603 & 8.18 & 0.36 & 7.53 & 0.19 \\
2459795.137166 & 8.48 & 0.43 & 6.87 & 0.22 \\
2459796.036095 & 6.42 & 0.37 & 5.92 & 0.19 \\
2459797.135164 & 6.35 & 0.33 & 6.11 & 0.17 \\
2459799.022427 & 6.74 & 0.35 & 5.97 & 0.17 \\
2459799.982018 & 5.32 & 0.34 & 4.95 & 0.17 \\
2459801.057614 & 6.38 & 0.36 & 4.96 & 0.19 \\
2459801.977320 & 5.88 & 0.45 & 4.31 & 0.25 \\
2459802.940267 & 5.93 & 0.42 & 4.61 & 0.21 \\
2459804.047114 & 4.75 & 0.36 & 4.50 & 0.18 \\
2459805.135009 & 3.35 & 0.32 & 2.70 & 0.16 \\
2459806.077454 & 2.94 & 0.31 & 2.42 & 0.16 \\
2459806.082119 & 2.48 & 0.31 & 2.26 & 0.16 \\
2459806.086788 & 2.83 & 0.32 & 2.62 & 0.16 \\
2459806.091449 & 2.68 & 0.32 & 2.11 & 0.16 \\
2459807.117948 & 1.39 & 0.31 & 1.20 & 0.15 \\
2460117.058774 & 3.43 & 0.35 & 2.61 & 0.19 \\
2460117.063462 & 3.43 & 0.35 & 3.07 & 0.18 \\
2460117.068140 & 3.15 & 0.35 & 2.90 & 0.18 \\
2460117.072813 & 4.08 & 0.34 & 3.42 & 0.18 \\
2460119.101419 & 2.72 & 0.31 & 2.31 & 0.17 \\
2460121.111875 & -0.17 & 0.31 & 0.13 & 0.17 \\
2460122.106457 & 1.54 & 0.31 & 2.21 & 0.17 \\
2460122.111554 & 0.85 & 0.32 & 1.43 & 0.17 \\
2460123.109925 & 1.78 & 0.31 & 1.85 & 0.17 \\
2460125.109304 & -0.00 & 0.30 & -0.13 & 0.17 \\
2460127.058743 & -1.23 & 0.33 & -1.79 & 0.18 \\
2460127.063513 & -1.02 & 0.33 & -1.89 & 0.18 \\
2460127.068045 & -1.33 & 0.32 & -1.69 & 0.17 \\
2460127.072750 & -0.65 & 0.31 & -1.08 & 0.17 \\
2460129.100610 & -1.66 & 0.32 & -2.00 & 0.17 \\
2460131.078938 & -2.57 & 0.34 & -2.93 & 0.17 \\
2460133.123639 & -3.40 & 0.32 & -3.10 & 0.17 \\
2460135.104334 & -5.04 & 0.36 & -4.79 & 0.19 \\
2460135.108853 & -4.67 & 0.34 & -5.21 & 0.18 \\
2460135.113438 & -4.37 & 0.35 & -4.95 & 0.18 \\
2460135.117939 & -4.47 & 0.37 & -5.38 & 0.19 \\
2460137.072109 & -5.73 & 0.34 & -5.81 & 0.18 \\
2460227.878406 & 16.56 & 0.33 & 15.14 & 0.15 \\
2460231.866414 & 14.44 & 0.31 & 14.00 & 0.14 \\
2460233.937023 & 14.08 & 0.34 & 12.77 & 0.16 \\
2460235.908700 & 14.00 & 0.31 & 13.06 & 0.14 \\
2460237.788609 & 13.19 & 0.33 & 11.55 & 0.16 \\
2460239.781390 & 12.23 & 0.33 & 10.42 & 0.15 \\
2460241.936769 & 10.91 & 0.36 & 9.30 & 0.17 \\
2460243.763092 & 10.41 & 0.35 & 9.10 & 0.16 \\
2460245.879571 & 8.70 & 0.35 & 7.06 & 0.16 \\
2460270.957594 & -3.98 & 0.39 & -5.45 & 0.21 \\
2460274.735803 & -7.75 & 0.35 & -9.75 & 0.18 \\
2460283.693922 & 0.75 & 0.33 & -0.07 & 0.18 \\
2460285.684591 & 11.34 & 0.33 & 11.11 & 0.18 \\
2460287.768267 & 15.61 & 0.42 & 14.87 & 0.24 \\
2460287.773770 & 14.34 & 0.40 & 13.06 & 0.23 \\
2460287.777636 & 15.55 & 0.37 & 14.32 & 0.21 \\
2460287.783013 & 15.64 & 0.40 & 14.77 & 0.23 \\
2460289.769822 & 16.20 & 0.34 & 15.93 & 0.19 \\
2460291.803387 & 17.20 & 0.33 & 16.13 & 0.19 \\
2460293.794083 & 15.78 & 0.33 & 15.85 & 0.19 \\
2460295.769727 & 16.44 & 0.33 & 16.10 & 0.19 \\
2460296.778448 & 15.28 & 0.34 & 14.49 & 0.19 \\
2460296.783032 & 15.45 & 0.34 & 14.75 & 0.19 \\
2460296.787125 & 17.09 & 0.34 & 16.52 & 0.19 \\
2460296.792445 & 15.28 & 0.34 & 15.04 & 0.19 \\
2460297.762720 & 14.77 & 0.37 & 13.73 & 0.20 \\
2460297.767874 & 13.06 & 0.36 & 13.10 & 0.19 \\
2460297.773058 & 14.63 & 0.36 & 14.45 & 0.19 \\
2460297.777701 & 14.72 & 0.36 & 14.51 & 0.19 \\
2460301.774260 & 12.25 & 0.48 & 11.01 & 0.28 \\
2460301.778449 & 14.44 & 0.47 & 13.43 & 0.27 \\
2460301.783011 & 12.11 & 0.38 & 11.21 & 0.22 \\
2460301.787862 & 13.36 & 0.49 & 12.05 & 0.28 \\
2460303.788686 & 12.04 & 0.34 & 11.80 & 0.19 \\
2460305.704028 & 11.60 & 0.35 & 11.46 & 0.20 \\
2460307.754244 & 10.64 & 0.35 & 10.09 & 0.20 \\
2460308.840262 & 9.67 & 0.36 & 9.37 & 0.20 \\
2460310.722774 & 8.48 & 0.35 & 7.94 & 0.19 \\
2460312.732028 & 8.36 & 0.35 & 7.42 & 0.20 
\label{table:HD3651-Data}
\end{longtable*}

\section{Dataset Parameters}

\begin{table}[]
    \centering
    \begin{tabular}{ccc}
        \hline
        \hline
        Parameter & Posterior & Prior \\
        \hline
        $\gamma_{\text{EXPRES}}\text{ [\,m\,s$^{-1}$]}$ & $-1.52^{+0.1}_{-0.09}$ & $\mathcal{U}[-10, 10]$ \\
        
        $\gamma_{\text{pre-NEID}}\text{ [\,m\,s$^{-1}$]}$ & $-4.16^{+0.14}_{-0.13}$ & $\mathcal{U}[-10, 10]$ \\
        
        $\gamma_{\text{post-NEID}}\text{ [\,m\,s$^{-1}$]}$ &  $-2.6^{+0.09}_{-0.09}$ & $ \mathcal{U}[-10, 10]$ \\
        
        $\gamma_{\text{pre-HIRES}}\text{ [\,m\,s$^{-1}$]}$ & $1.75^{+0.63}_{-0.64}$ & $\mathcal{U}[-10, 10]$ \\
        
        $\gamma_{\text{post-HIRES}}\text{ [\,m\,s$^{-1}$]}$ & $-2.57^{+0.28}_{-0.29}$ & $\mathcal{U}[-10, 10]$ \\
        
        $\gamma_{\text{MX Run10 Red}}\text{ [\,m\,s$^{-1}$]}$ & $-2.14^{+0.13}_{-0.12}$ & $\mathcal{U}[-30, 30]$ \\
        
        $\gamma_{\text{MX Run10 Blue}}\text{ [\,m\,s$^{-1}$]}$ & $-1.23^{+0.08}_{-0.08}$ & $\mathcal{U}[-30, 30]$ \\
        
        $\gamma_{\text{MX Run10 - Run6 Red}}\text{ [\,m\,s$^{-1}$]}$ & $-8.42^{+0.19}_{-0.19}$ & $\mathcal{N}[-8.54,0.71]$ \\
        $\gamma_{\text{MX Run10 - Run6 Blue}}\text{ [\,m\,s$^{-1}$]}$ & $-5.84^{+0.16}_{-0.17}$ & $\mathcal{N}[-5.61,0.71]$ \\
        
        $\gamma_{\text{MX Run10 - Run7 Red}}\text{ [\,m\,s$^{-1}$]}$ & $-6.78^{+0.15}_{-0.15}$ & $\mathcal{N}[-6.52,0.71]$ \\
        $\gamma_{\text{MX Run10 - Run7 Blue}}\text{ [\,m\,s$^{-1}$]}$ & $-4.81^{+0.19}_{-0.18}$ & $\mathcal{N}[-3.74,0.71]$ \\
        
        $\gamma_{\text{MX Run10 - Run9 Red}}\text{ [\,m\,s$^{-1}$]}$ & $-2.5^{+0.3}_{-0.28}$ & $\mathcal{N}[-1.61,0.72]$ \\
        $\gamma_{\text{MX Run10 - Run9 Blue}}\text{ [\,m\,s$^{-1}$]}$ & $-1.73^{+0.24}_{-0.22}$ & $\mathcal{N}[-0.90,0.71]$ \\
        
        $\gamma_{\text{MX Run10 - Run11 Red}}\text{ [\,m\,s$^{-1}$]}$ & $0.09^{+0.16}_{-0.16}$ & $\mathcal{N}[-0.23,0.71]$ \\
        $\gamma_{\text{MXRun10 -  Run11 Blue}}\text{ [\,m\,s$^{-1}$]}$ & $0.64^{+0.16}_{-0.16}$ & $\mathcal{N}[0.16,0.71]$ \\
        
        $\gamma_{\text{MX Run10 - Run12 Red}}\text{ [\,m\,s$^{-1}$]}$ & $8.06^{+0.2}_{-0.2}$ & $\mathcal{N}[7.73,0.77]$ \\
        $\gamma_{\text{MX Run10 - Run12 Blue}}\text{ [\,m\,s$^{-1}$]}$ & $7.56^{+0.22}_{-0.21}$ & $\mathcal{N}[7.95,0.71]$ \\
        
        $\gamma_{\text{MX Run10 - Run13 Red}}\text{ [\,m\,s$^{-1}$]}$ & $9.09^{+0.35}_{-0.4}$ & $\mathcal{N}[8.65,0.71]$ \\
        $\gamma_{\text{MX Run10 - Run13 Blue}}\text{ [\,m\,s$^{-1}$]}$ & $8.33^{+0.23}_{-0.24}$ & $\mathcal{N}[8.39,0.71]$ \\
        
        $\gamma_{\text{MX Run10 - Run14 Red}}\text{ [\,m\,s$^{-1}$]}$ & $8.94^{+0.18}_{-0.18}$ & $\mathcal{N}[9.06,0.71]$ \\
        $\gamma_{\text{MX Run10 - Run14 Blue}}\text{ [\,m\,s$^{-1}$]}$ & $9.07^{+0.19}_{-0.18}$ & $\mathcal{N}[8.58,0.74]$ \\
        
        $\sigma_{\text{EXPRES}}\text{ [\,m\,s$^{-1}$]}$ & $0.68^{+0.07}_{-0.06}$ & $\log \mathcal{U}[10^{-1}, 5]$ \\
        
        $\sigma_{\text{pre-NEID}}\text{ [\,m\,s$^{-1}$]}$ & $0.79^{+0.11}_{-0.1}$ & $\log \mathcal{U}[10^{-1}, 5]$ \\
        
        $\sigma_{\text{post-NEID}}\text{ [\,m\,s$^{-1}$]}$ & $0.95^{+0.08}_{-0.07}$ & $\log \mathcal{U}[10^{-1}, 5]$ \\
        
        $\sigma_{\text{pre-HIRES}}\text{ [\,m\,s$^{-1}$]}$ & $3.56^{+0.53}_{-0.46}$ & $\log \mathcal{U}[10^{-1}, 5]$ \\
        
        $\sigma_{\text{post-HIRES}}\text{ [\,m\,s$^{-1}$]}$ & $3.28^{+0.21}_{-0.19}$ & $\log \mathcal{U}[10^{-1}, 5]$ \\
        
        $\sigma_{\text{MX Run6 Red}}\text{ [\,m\,s$^{-1}$]}$ & $0.23^{+0.18}_{-0.1}$ & $\log \mathcal{U}[10^{-1}, 5]$ \\
        $\sigma_{\text{MX Run6 Blue}}\text{ [\,m\,s$^{-1}$]}$ & $0.3^{+0.15}_{-0.1}$ & $\log \mathcal{U}[10^{-1}, 5]$ \\
        
        $\sigma_{\text{MX Run7 Red}}\text{ [\,m\,s$^{-1}$]}$ & $0.33^{+0.14}_{-0.13}$ & $\log \mathcal{U}[10^{-1}, 5]$ \\
        $\sigma_{\text{MX Run7 Blue}}\text{ [\,m\,s$^{-1}$]}$ & $0.75^{+0.14}_{-0.11}$ & $\log \mathcal{U}[10^{-1}, 5]$ \\
        
        $\sigma_{\text{MX Run9 Red}}\text{ [\,m\,s$^{-1}$]}$ & $0.48^{+0.36}_{-0.24}$ & $\log \mathcal{U}[10^{-1}, 5]$ \\
        $\sigma_{\text{MX Run9 Blue}}\text{ [\,m\,s$^{-1}$]}$ & $0.43^{+0.27}_{-0.16}$ & $\log \mathcal{U}[10^{-1}, 5]$ \\
        
        $\sigma_{\text{MX Run10 Red}}\text{ [\,m\,s$^{-1}$]}$ & $0.62^{+0.1}_{-0.09}$ & $\log \mathcal{U}[10^{-1}, 5]$ \\
        $\sigma_{\text{MX Run10 Blue}}\text{ [\,m\,s$^{-1}$]}$ & $0.6^{+0.08}_{-0.07}$ & $\log \mathcal{U}[10^{-1}, 5]$ \\
        
        $\sigma_{\text{MX Run11 Red}}\text{ [\,m\,s$^{-1}$]}$ & $0.44^{+0.11}_{-0.09}$ & $\log \mathcal{U}[10^{-1}, 5]$ \\
        $\sigma_{\text{MX Run11 Blue}}\text{ [\,m\,s$^{-1}$]}$ & $0.59^{+0.11}_{-0.08}$ & $\log \mathcal{U}[10^{-1}, 5]$ \\
        
        $\sigma_{\text{MX Run12 Red}}\text{ [\,m\,s$^{-1}$]}$ & $0.36^{+0.19}_{-0.15}$ & $\log \mathcal{U}[10^{-1}, 5]$ \\
        $\sigma_{\text{MX Run12 Blue}}\text{ [\,m\,s$^{-1}$]}$ & $0.61^{+0.2}_{-0.14}$ & $\log \mathcal{U}[10^{-1}, 5]$ \\
        
        $\sigma_{\text{MX Run13 Red}}\text{ [\,m\,s$^{-1}$]}$ & $0.34^{+0.56}_{-0.19}$ & $\log \mathcal{U}[10^{-1}, 5]$ \\
        $\sigma_{\text{MX Run13 Blue}}\text{ [\,m\,s$^{-1}$]}$ &  $0.22^{+0.26}_{-0.09}$ & $\log \mathcal{U}[10^{-1}, 5]$ \\
        
        $\sigma_{\text{MX Run14 Red}}\text{ [\,m\,s$^{-1}$]}$ & $0.71^{+0.13}_{-0.11}$ & $\log \mathcal{U}[10^{-1}, 5]$ \\
        $\sigma_{\text{MX Run14 Blue}}\text{ [\,m\,s$^{-1}$]}$ & $0.86^{+0.14}_{-0.11}$ & $\log \mathcal{U}[10^{-1}, 5]$ \\
        \hline
    \end{tabular}
    \caption{This table summarizes the posteriors and the priors for derived dataset parameters.}
    \label{tab:dataset-priors}
\end{table}

\bibliography{sample631}{}
\bibliographystyle{aasjournal}



\end{document}